\documentclass[prc,aps,showpacs]{revtex4}
\usepackage{graphicx}
\usepackage{subfig}
\usepackage{float}
\begin{document}

\title{Testing the spin-cutoff parameterization with shell-model calculations}

\author{William M. Spinella}
\author{Calvin W. Johnson}
\affiliation{Department of Physics, San Diego State University,
5500 Campanile Drive, San Diego, CA 92182-1233}
\affiliation{Computational Science Research Center, San Diego State University,
5500 Campanile Drive, San Diego, CA 92182-1245}


\begin{abstract}
The nuclear level density, an important input to Hauser-Feshbach 
calculations, depends not only on excitation energy but also 
on angular momentum $J$.  The $J$-dependence of the level density 
at fixed excitation energy $E_x$ is usually parameterized via the
spin-cutoff factor $\sigma$. We carefully test the statistical accuracy 
of this parameterization for a large number of spectra computed 
using semi-realistic interactions in the interacting shell model, 
with a nonlinear least-squares fit of $\sigma$ and 
finding the error bar in $\sigma$.  
The spin-cutoff parameterization works well as long as there are enough 
states to be statistical. In turn, the spin-cutoff factor 
can be related to the average value of $J^2$ at a fixed 
excitation energy, and we briefly investigate extracting 
$\langle \hat{J}^2 (E_x)\rangle$ from a thermal calculation 
such as one might do via Monte Carlo.
\end{abstract}

\maketitle

\section{Introduction}

An important input into compound nuclear reactions such as statistical neutron capture is the level density\cite{HF52}.  
 The level density  is deceptively simple: 
the number of levels 
per unit of excitation energy.  But when thousands or millions of states or more are involved, as  in the statistical regime, 
experimental measurement becomes challenging and theoretical calculations are also difficult if not intractable. 
 Because of this the level density is one of the most uncertain inputs 
into reaction calculations \cite{RTK97}.

(Note: one should be careful to distinguish between 
the \textit{level density}, which does not include the $2J+1$ degeneracy in $M$, and  the \textit{state density}  which does include
 the $2J+1$ degeneracy in $M$.)

For calculation of statistical neutron capture rates one 
needs not only the density of levels $\rho$ as a function of excitation energy $E_x$ but also as a function of 
parity and angular momentum $J$.
While there are different parameterizations for the level density $\rho(E_x)$, such as the 
back-shifted Fermi-gas \cite{Bet36,Gil65,Boh69} or constant temperature \cite{Gil65}, for the dependence on $J$ the parameterization  \cite{Eri60} 
\begin{equation}
\rho_J(E_x) = \rho(E_x) f(J), \,\,\, f(J) = \frac{2J+1}{2\sigma^2} \exp \left ( -J(J+1)/ 2\sigma^2 \right ),
\label{Ericson}
\end{equation}
is universally used,
where $\sigma^2$ is called the \textit{spin-cutoff factor}.  (This normalization is for 
the level density; for the state density the spin-cutoff parameterization  has a different normalization, 
found below in Eq.~(\ref{Ericson_state}).)

A number of other works have investigated the spin distribution of levels, both experimentally 
\cite{Gri74,Gri78,vEB08, vEB09} 
and theoretically \cite{Gri83,Hua00,ALN07,VRHA09}. While these papers have compared the actual spin distribution against the 
spin-cutoff parameterization,  in this paper we focus on \textit{quantifying} 
the statistical goodness of the spin-cutoff parameterization  by computing the error bars on the spin-cutoff factor.

Towards this end, we have generated a  number of theoretical nuclear spectra, as described in more 
detail in Section \ref{SM}.  We then put the data into energy bins and, in each energy bin, did a nonlinear 
least-squares fit for $\sigma$ using Eq.~(\ref{Ericson}), including deriving the error bars on $\sigma$.  
The fitting methodology and results we give  in Section \ref{results}.

In general we find the unsuprising but gratifying result that the spin-cutoff parameterization does a good job of describing 
the spin distribution, and that the error bars on $\sigma$ decrease as the number of levels in a bin increase.
If Eq.~(\ref{Ericson}) were wrong, the error bars would not decrease systematically. 

Because Eq.~(\ref{Ericson}) has only the one parameter $\sigma$, it is tempting to try to extract the spin-cutoff factor 
directly rather than fitting it, for example relating  $\sigma^2$ to the average value of $J(J+1)$ at a given excitation 
energy. 
 This would be useful in particular for Monte Carlo calculations such as the so-called shell-model 
Monte Carlo (SMMC) \cite{Lang93,KDL97} from which state densities can be extracted 
\cite{ALN07, VRHA09, Orm97,NA97,NA98,LA01,ABFL05,AFN08}; 
although a $J$-projected versions of SMMC has been 
developed \cite{ALN07, VRHA09}, it is  computationally very
intensive. By contrast, it is relatively easy to compute thermally-averaged expectation values 
of operators such as $\hat{J}^2$ in SMMC.  
 Because one needs the average as a function of excitation energy and not temperature, one needs to 
extract the energy-averaged $\langle J(J+1)(E) \rangle $ from thermally-weighted values, in the same way one inverts the 
Laplace transform to extract the state density from the thermal partition function.  This we discuss  in Section \ref{QMC}.

\section{Shell-model framework}
\label{SM}

We work in the framework of the configuration-interaction shell model \cite{BG77,BW88,ca05}. Here one defines a finite 
single-particle space and has as input single-particle energies and two-body matrix elements. We use 
semi-realistic/semi-phenomenological matrix elements, meaning each interaction starts from realistic fits 
to scattering data, is then renormalized, and finally tuned to reproduce binding energies and 
spectra for a specified mass region (see \cite{BG77,USDB} for details of methodology). 

  All model spaces assume some 
inert core and valence particles.
The three model spaces we work in and their interactions are:

\noindent $\bullet$ \textit{sd}, or the 1$s_{1/2}$-$0d_{3/2}$-$0d_{5/2}$ valence space, 
assuming an inert $^{16}$O core; the interaction is the universal $sd$-interaction `B,'
or USDB \cite{USDB};

\noindent $\bullet$ \textit{pf}, or the $1p_{1/2}$-$1p_{3/2}$-$0f_{5/2}$-$0f_{7/2}$ valence space, 
assuming an inert $^{40}$Ca core; the interaction is the monopole-modified Kuo-Brown G-matrix interaction 
version 3G, or KB3G \cite{KB68,Pov01};

\noindent $\bullet$ and finally the \textit{ps-d}$_{5/2}$, or $0p_{3/2}$-$0p_{1/2}$-$0d_{5/2}$-
$1s_{1/2}$ valence space, assuming an inert $^4$He core; the interaction is a 
hybrid of  Cohen-Kurath (CK) matrix elements in
the $0p$ shell\cite{CK65}, the older universal $sd$ interaction of Wildenthal \cite{Wildenthal} in
the $0d_{5/2}$-$1s_{1/2}$ space, and the Millener-Kurath (MK)
$p$-$sd$ cross-shell matrix elements \cite{MK75}.  
We leave out the $0d_{3/2}$ orbit to make 
calculations tractable.  Within the $p$ and
$sd$ spaces we use the original spacing of the single-particle
energies for the CK and Wildenthal interactions, respectively, but then
shift the $sd$ single-particle energies  relative to the
$p$-shell single particle energies to place the first $3^-$ state in $^{16}$O at approximately
$6.1$ MeV above the ground state. The rest of the spectrum, in
particular the first excited $0^+$ state, is not very good, but the
idea is to have a non-trivial model, not exact reproduction of the
spectrum. This model space and interaction allows us to consider model nuclei with 
both parities and to investigate parity-mixing in the HF state.

Finally, we computed the many-body spectra using the BIGSTICK configuration-interaction 
diagonalization code \cite{BIGSTICK}.  Table \ref{tab:list} lists the nuclides computed 
and the single-particle
spaces used. In this paper we display a representative sample, but graphs for all nuclides 
in Table \ref{tab:list} are included in the supplemental material \cite{SJsuppl}.

\begin{table}
	\centering
	\caption{Table of analyzed nuclides.}
	\begin{tabular}{|c|c|c|}
		\hline
		{\textit{sd}} & {\textit{pf}} & {\textit{p-sd}$_{5/2}$}\\
\hline
$^{22 -27}$Na    & $^{46-52}$Ca       &  $^{11-13}$C\\
$^{24 -29}$Mg   &  $^{45-47,49}$Sc  &   $^{14}$N\\
$^{26- 30}$Al   &  $^{44-47}$Ti         &  $^{16-17}$O\\
$^{28-32}$Si    &   $^{46}$V               &   $^{20}$Ne\\
$^{30-33}$P   & & \\
$^{32-34}$S     &   &     \\
$^{34 -35}$Cl   &  &    \\
	\hline
	\end{tabular}
	\label{tab:list}
\end{table}

\section{Results}
\label{results}

To analyze the spin distribution of shell-model spectra, we binned the data into 1 MeV bins (our results were 
insensitive to the size of the bins, as we discuss at the end of this section), 
and did a non-linear least-squares fit to the Ericson formula 
using the MPFIT code \cite{Mar09,MPFIT}, an implementation of the Levenberg-Marquardt algorithm 
\cite{Mor78}.    
 The error bar on the  least-squares-fit 
spin-cutoff parameter 
is just the standard square root of the diagonal of the covariance matrix, which in our case was 
 a trivial $1\times 1$ matrix.
For the purpose of the fit we estimate the error of each $J$-projected density $\rho_J$ in an
energy bin to be the standard counting statistical error, $\Delta \rho_J = \sqrt{\rho_{J}}$; bins
with $\rho_{J}=0$ are assigned an error of 1.   
Our initial $\chi^2$ per degree of freedom, or reduced $\chi^2$,
\begin{equation}
\chi^2_\mathrm{red} = \frac{1}{N_J -1} \sum_J \left (\rho_J - \rho f(J) \right )^2/ \Delta^2\rho_J,
\end{equation}
nearly always fell below 1 and frequently $<<1$.  Because we have only one parameter, overfitting is unlikely; instead we 
suspect the errors are non-Gaussian, asymmetric (for example, no density can be less than zero) 
and/or correlated.  To obtain a reduced $\chi^2=1$ for a given energy bin we uniformly scaled the 
errors $\Delta \rho_J$ in an energy bin, in almost all cases making them smaller.

(To understand in detail the true nature of our errors is beyond the scope of this investigation.
Inspired by our observation, however, that reducing the errors led to reduced $\chi^2=1$, we tried 
 a different model for errors, namely $\Delta \rho_J = \rho_J^{1/4}$. The resulting 
reduced $\chi^2$ were scattered about 
1, but the best-fit spin-cutoff factors and uncertainties in $\sigma$ changed by less than $5\%$. 
Therefore, although we have some ambiguities in understanding our uncertainties, we stand behind 
 our conclusion, that the Ericson parameterization of $\rho_J$ is statistically a 
very good model.)

We analyzed spectra from a wide range of nuclides in three different model spaces, and considered 
even-even, odd-odd, and odd-$A$ nuclei.  We will show representative plots from a number of different 
calculations; additional graphs can be found in the supplemental material \cite{SJsuppl}.

\begin{figure}
\subfloat{
\includegraphics[width=0.4\textwidth]{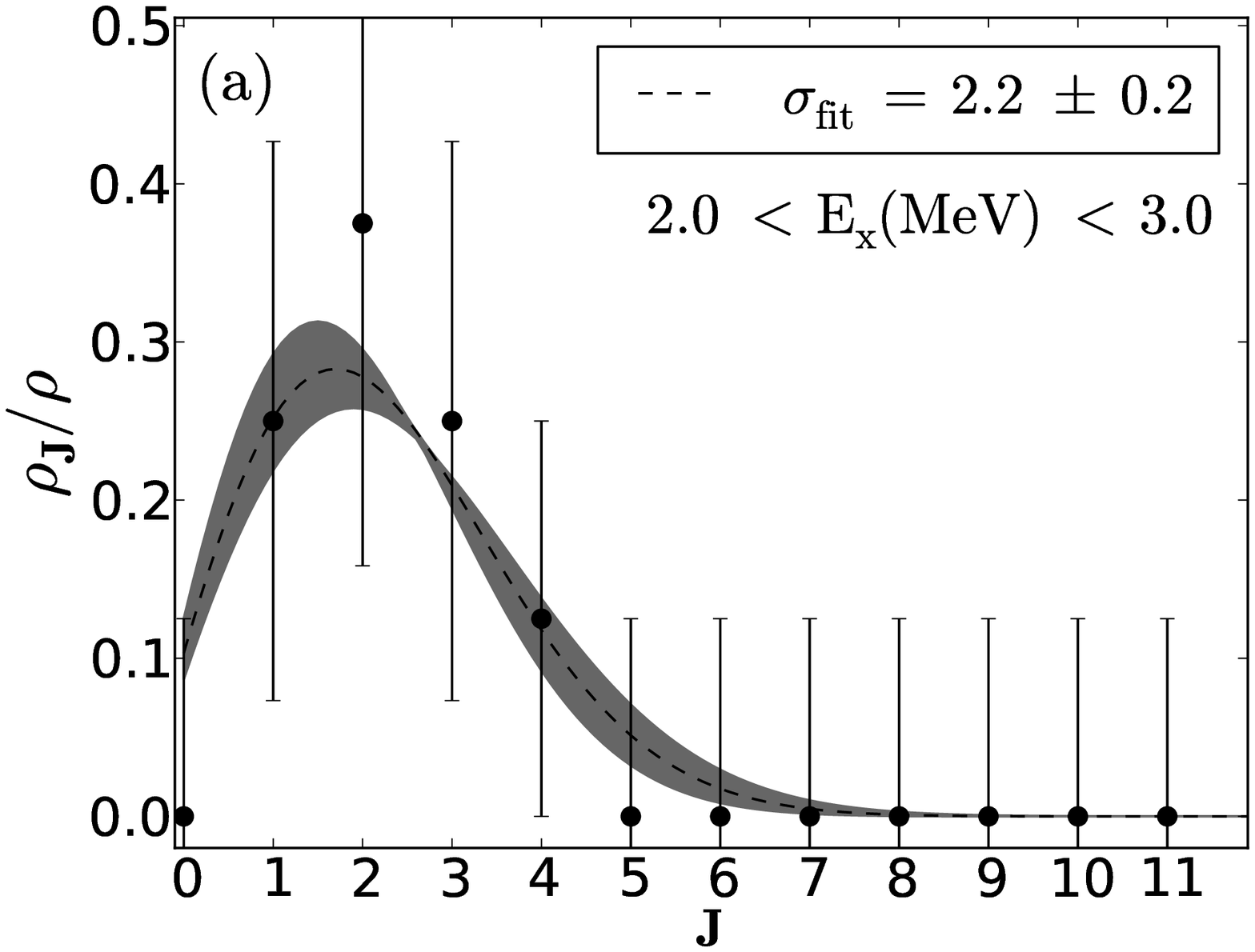}
}
\subfloat{
\includegraphics[width=0.4\textwidth]{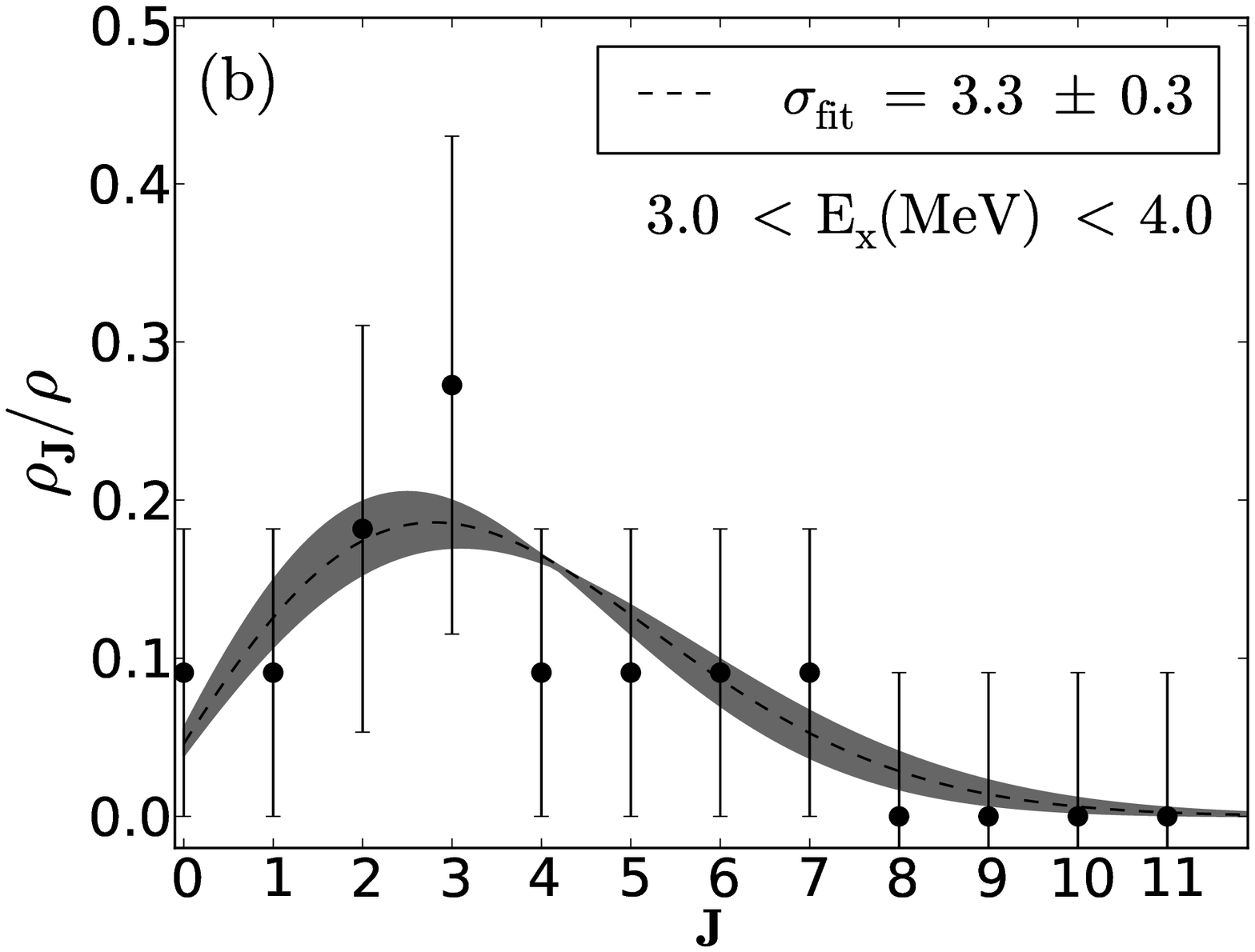}
} \\

\subfloat{
\includegraphics[width=0.4\textwidth]{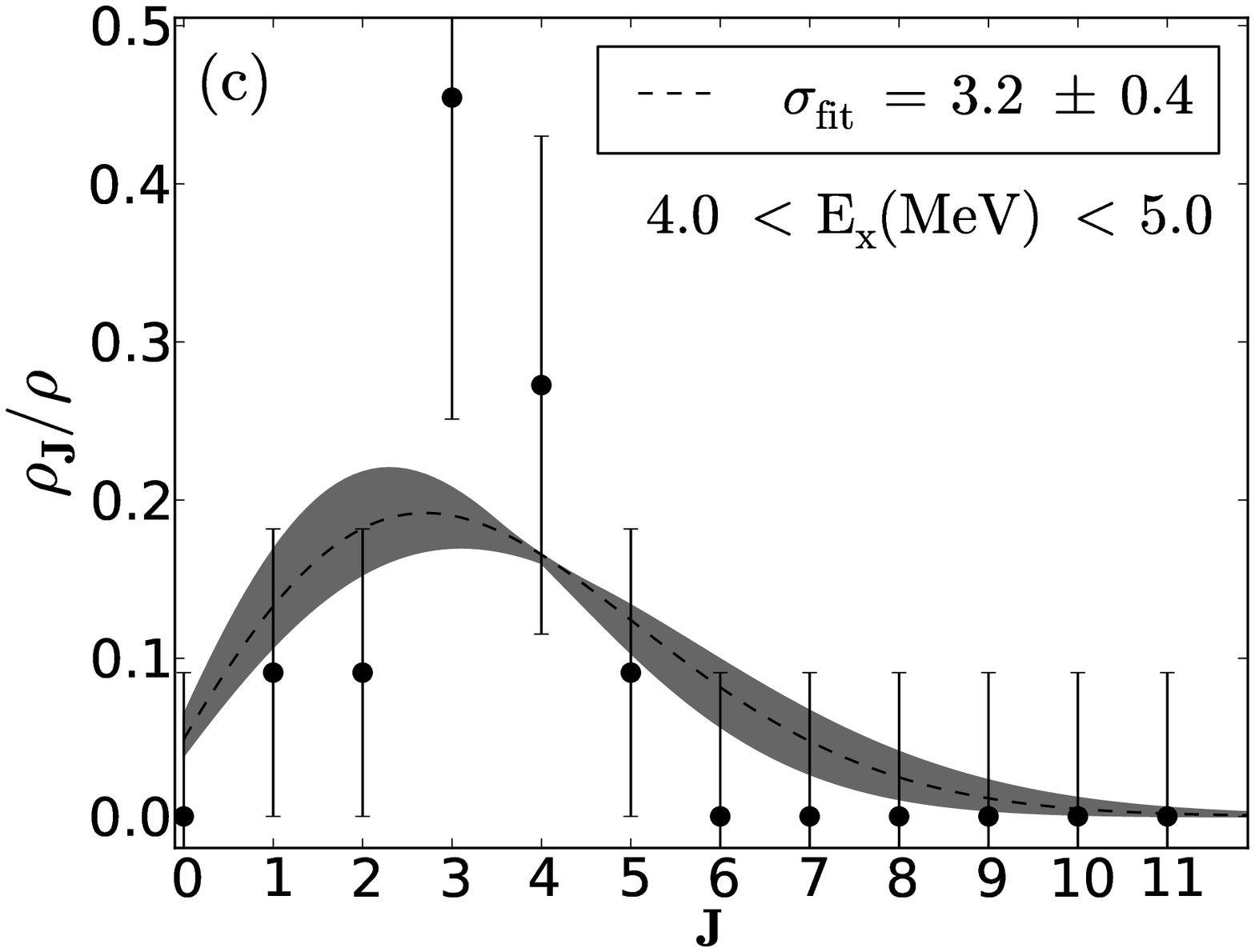}
}
\subfloat{
\includegraphics[width=0.4\textwidth]{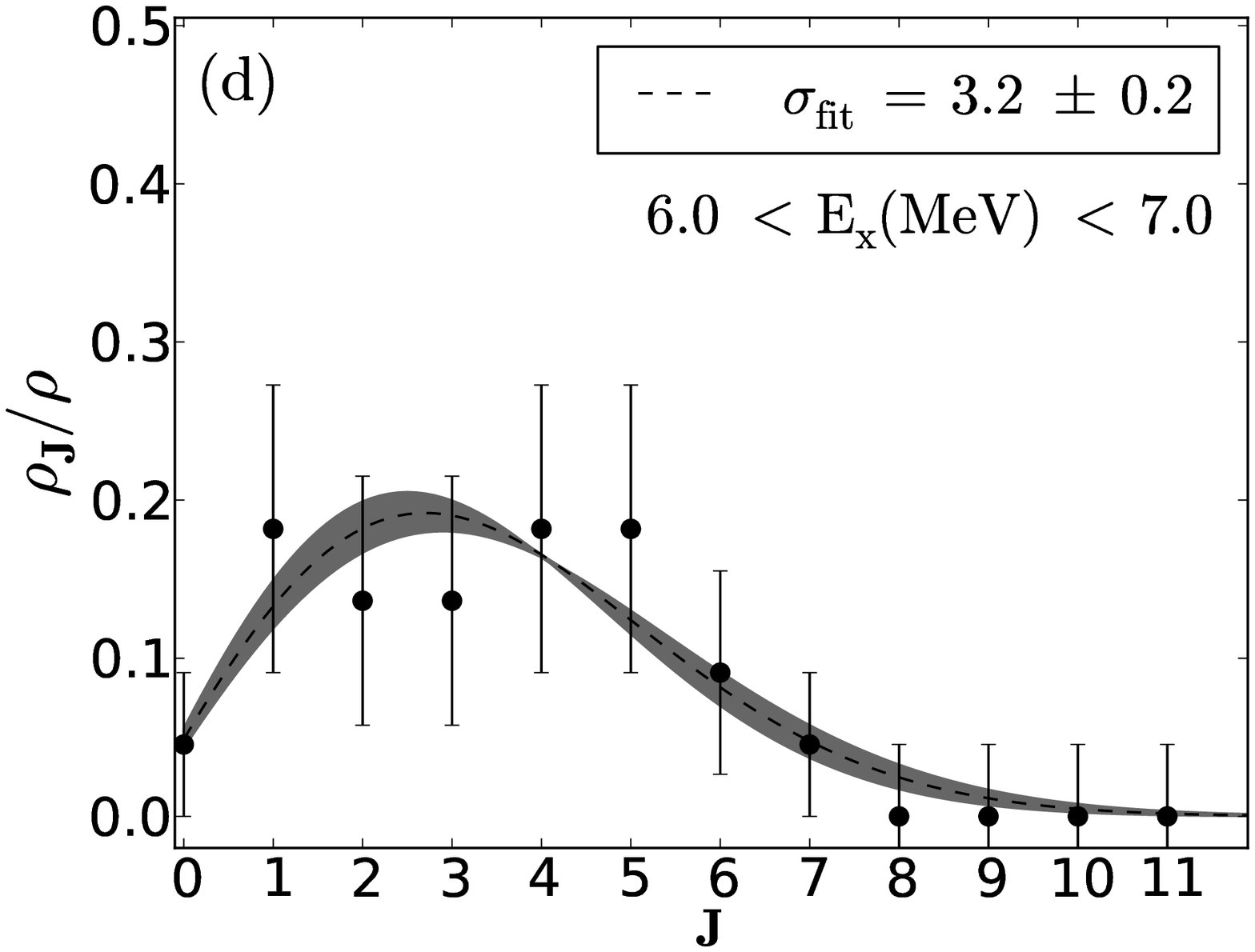}
} \\
\subfloat{
\includegraphics[width=0.4\textwidth]{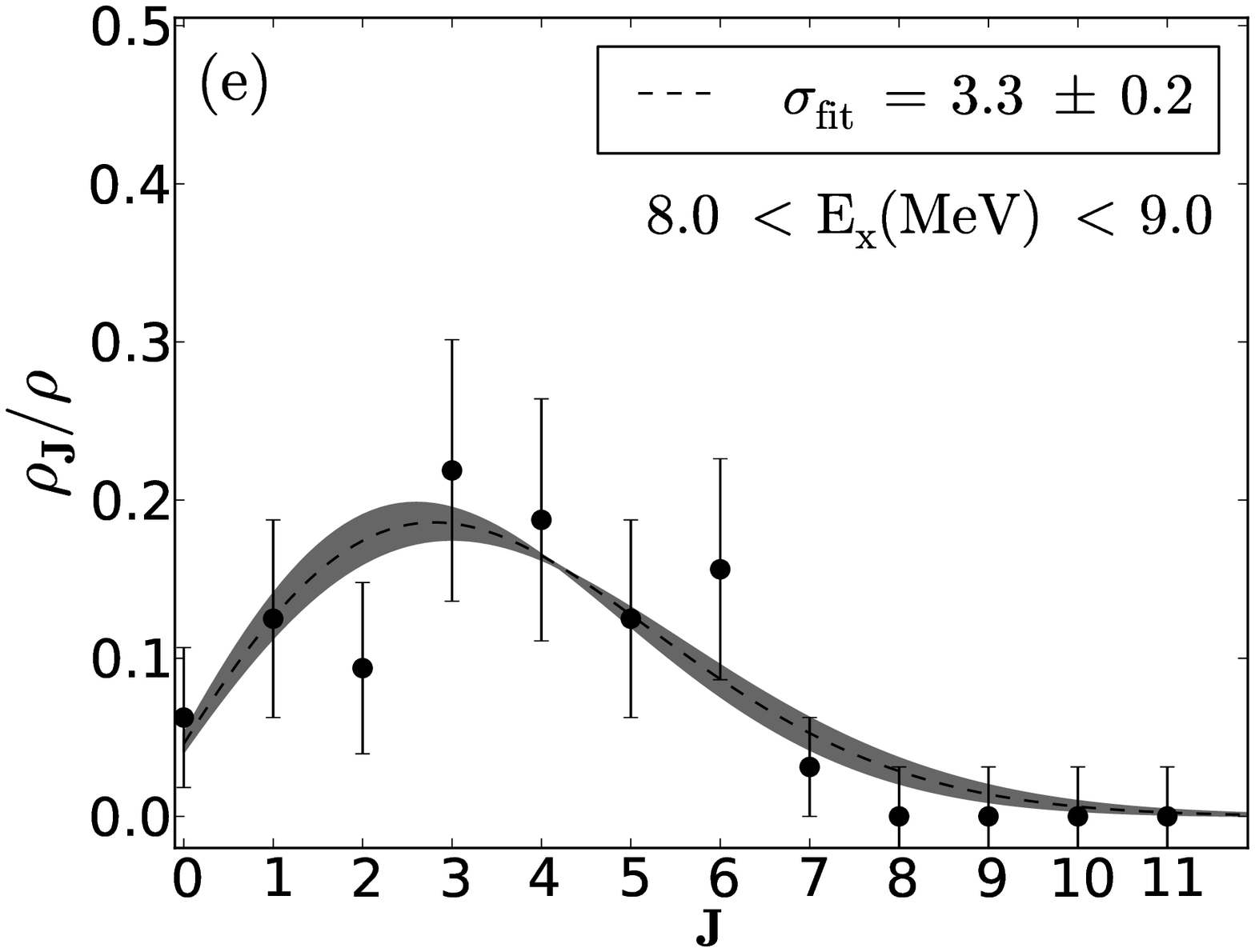}
}
\subfloat{
\includegraphics[width=0.4\textwidth]{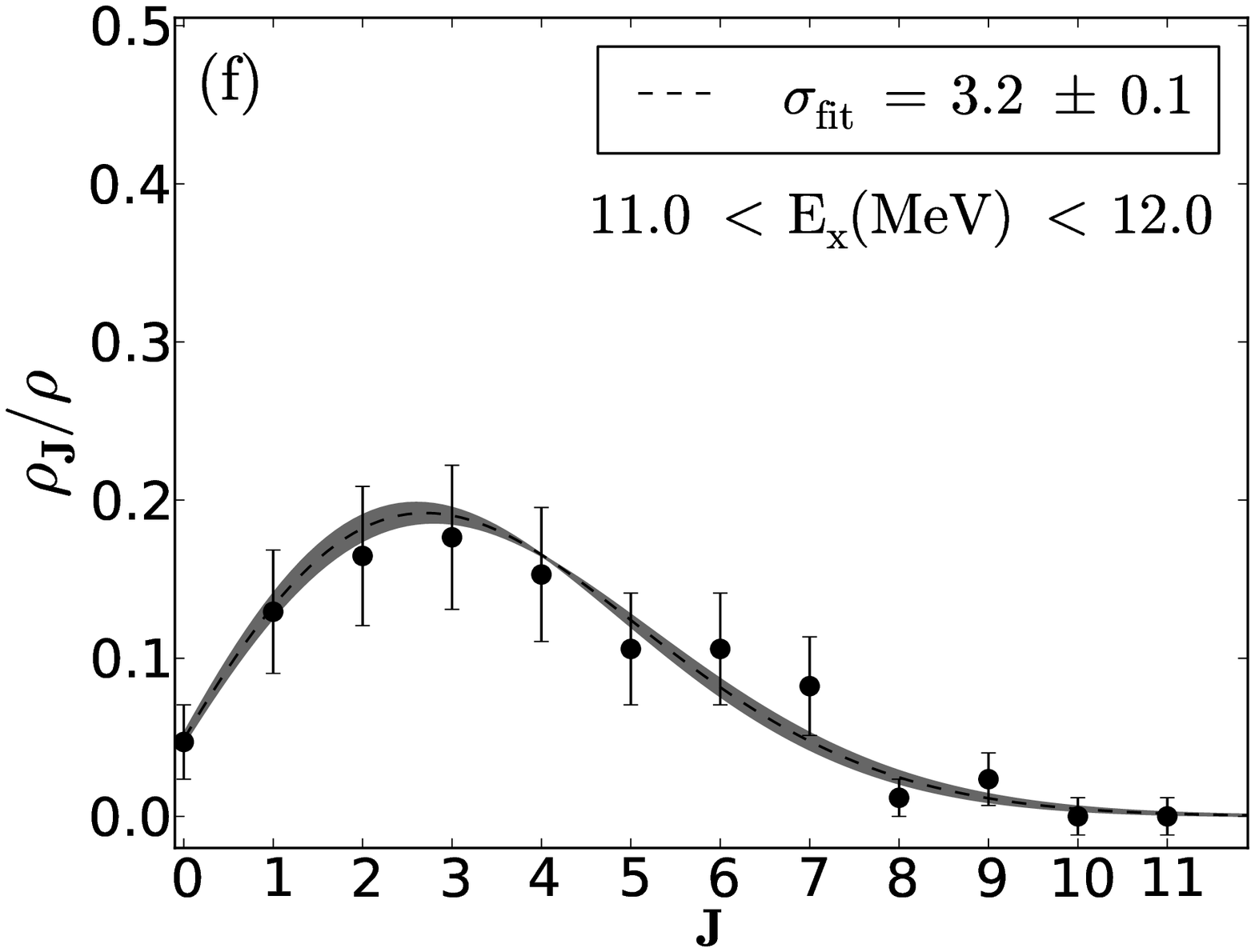}
}
\caption{$^{26}$Al in the $sd$ shell. The error bars are standard $\sqrt{\rho_J}$ 
statistical error in counting the number of states in a given bin of energy 1.0 MeV and 
fixed $J$. The dashed line is the continuous spin distribution using $\sigma_\mathrm{fit}$ and 
the gray shading is the one-standard-deviation envelope.  \label{Al26}}
\end{figure}

\begin{figure}

\subfloat{
\includegraphics[width=0.4\textwidth]{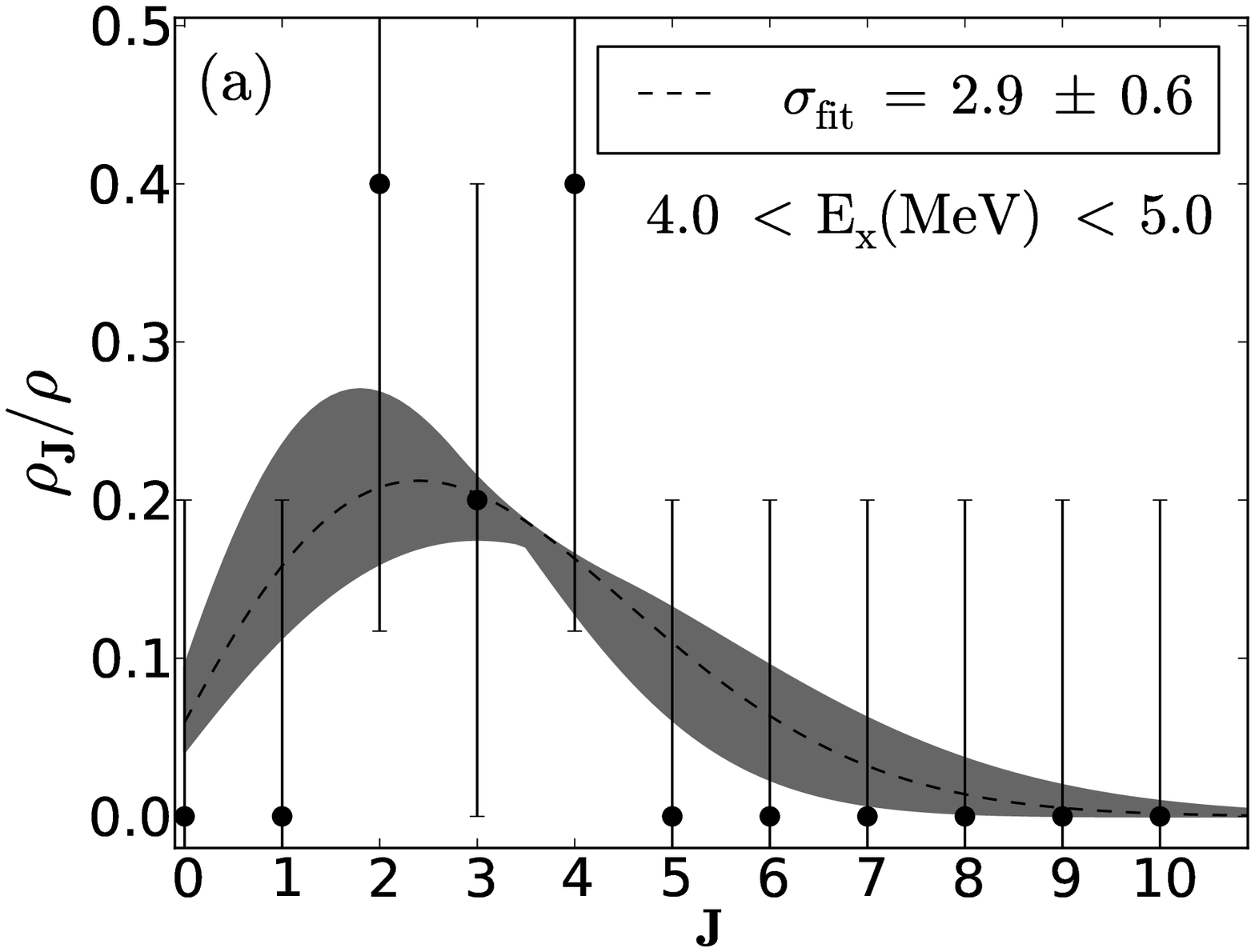}
}
\subfloat{
\includegraphics[width=0.4\textwidth]{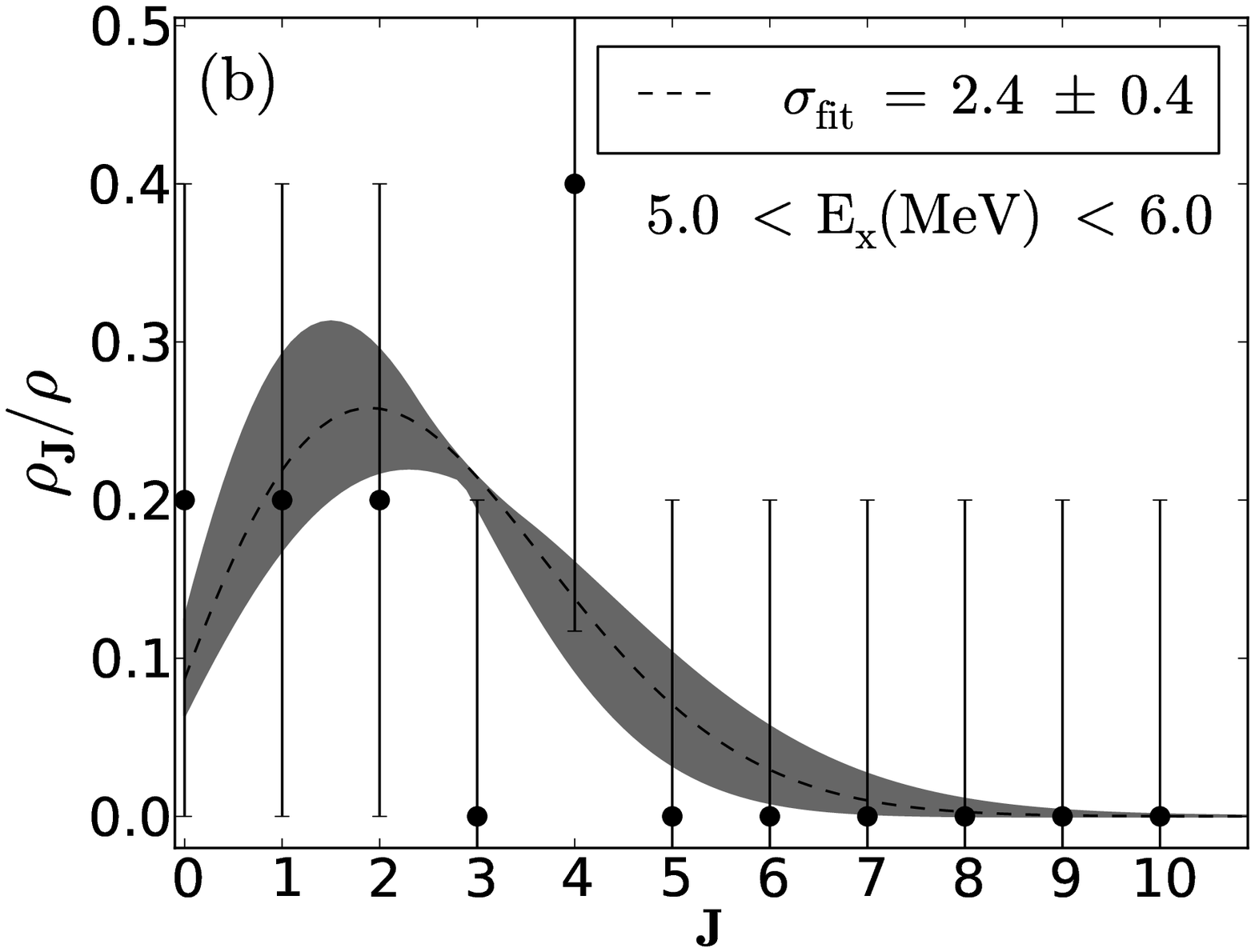}
} \\
\subfloat{
\includegraphics[width=0.4\textwidth]{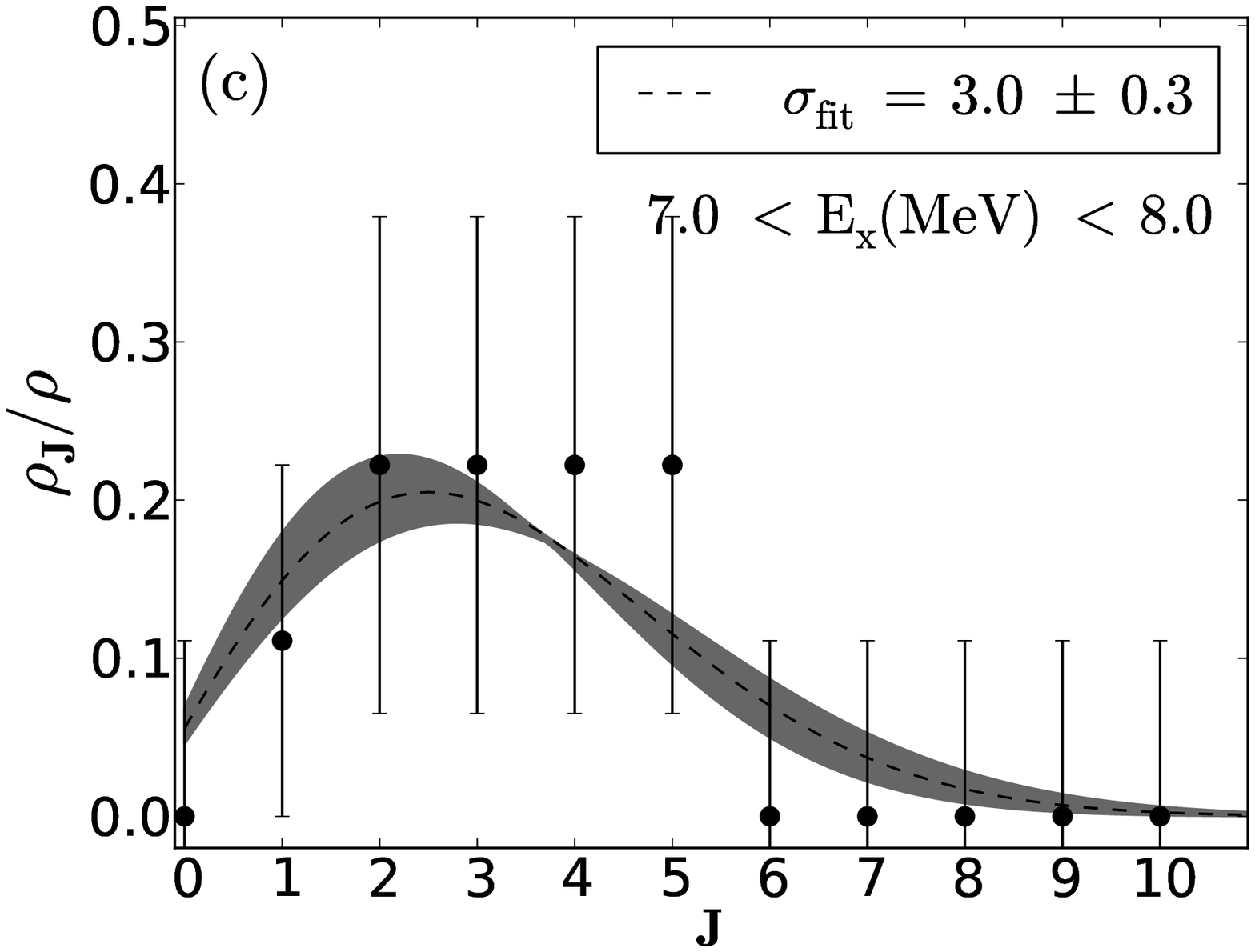}
}
\subfloat{
\includegraphics[width=0.4\textwidth]{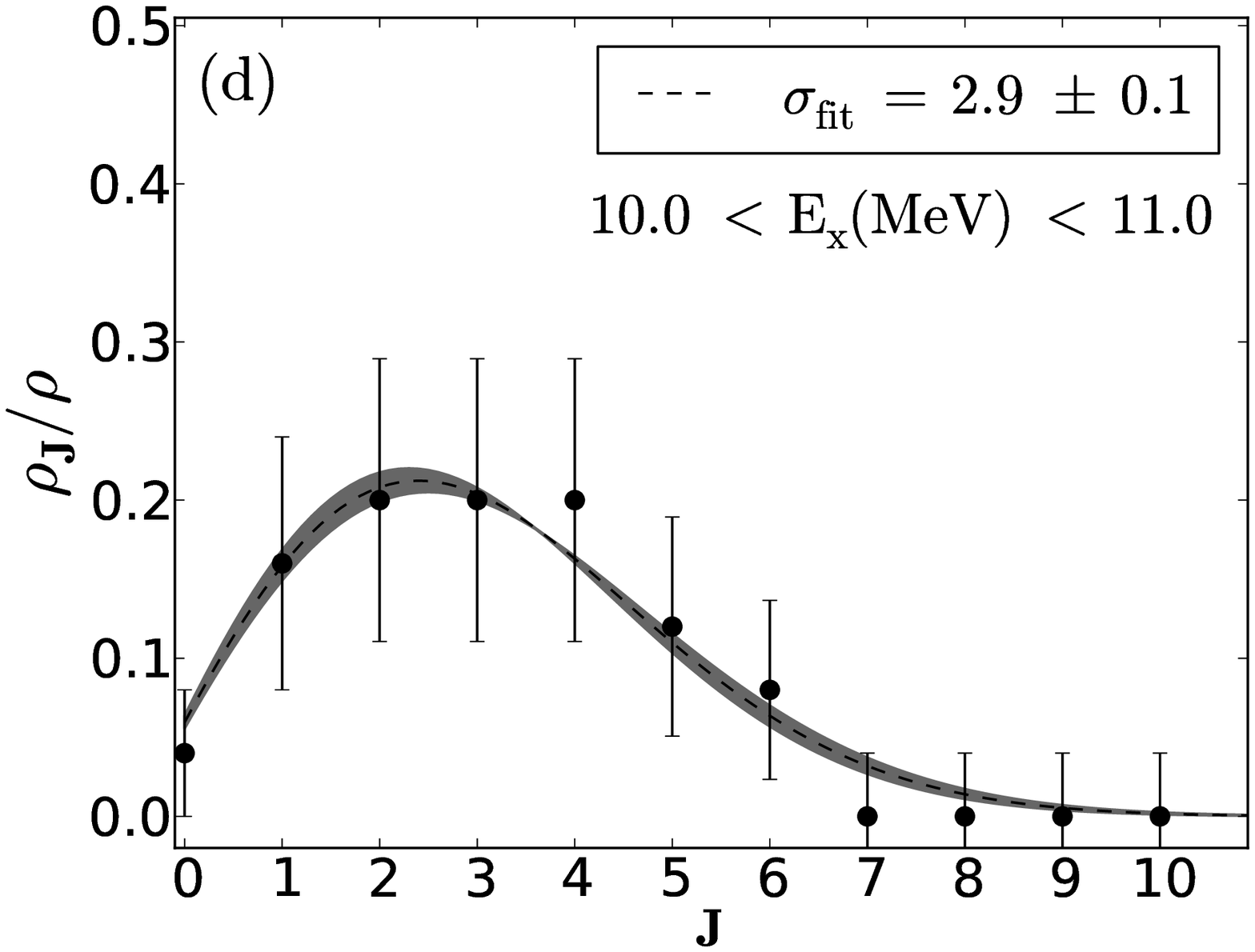}
} \\
\subfloat{
\includegraphics[width=0.4\textwidth]{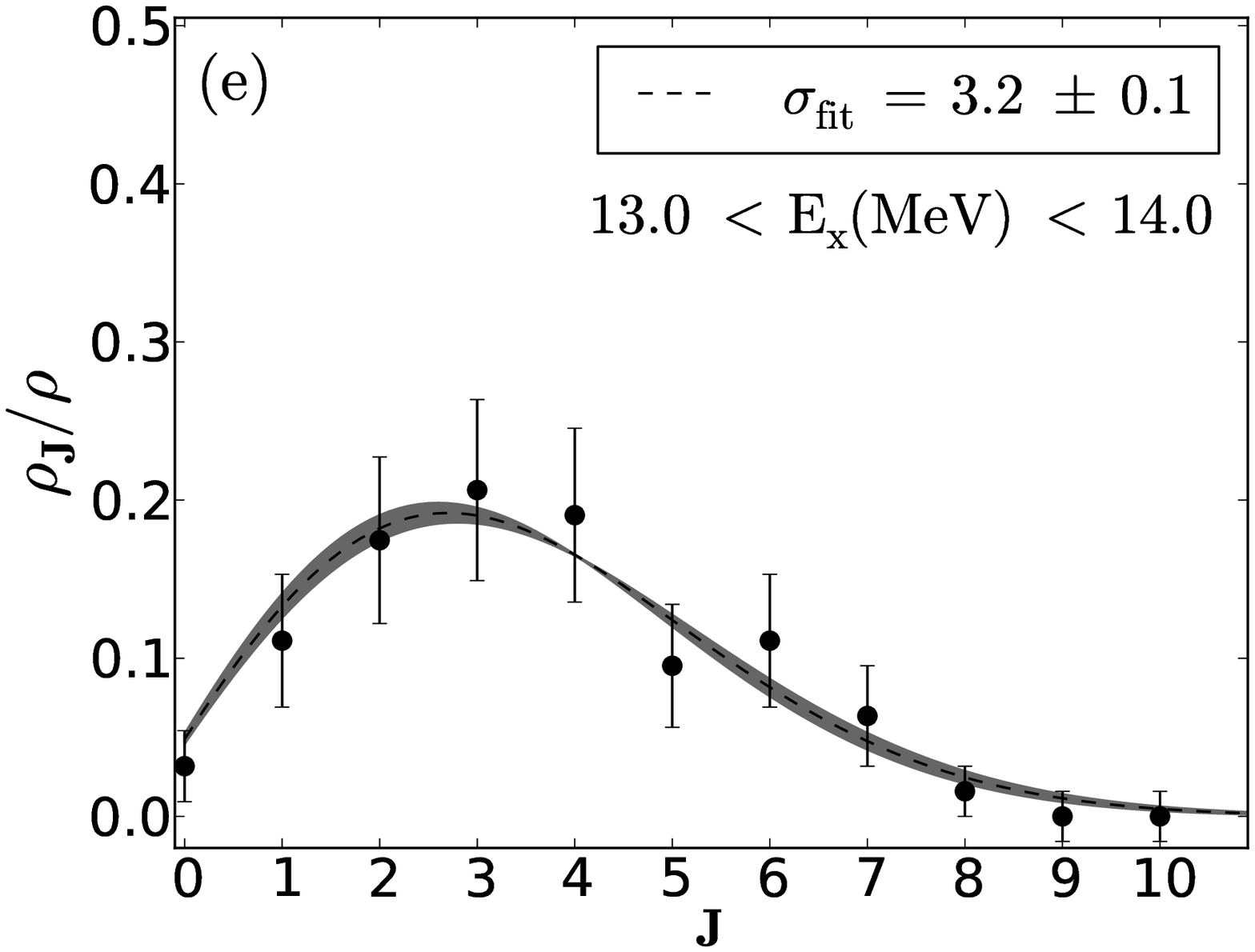}
}
\subfloat{
\includegraphics[width=0.4\textwidth]{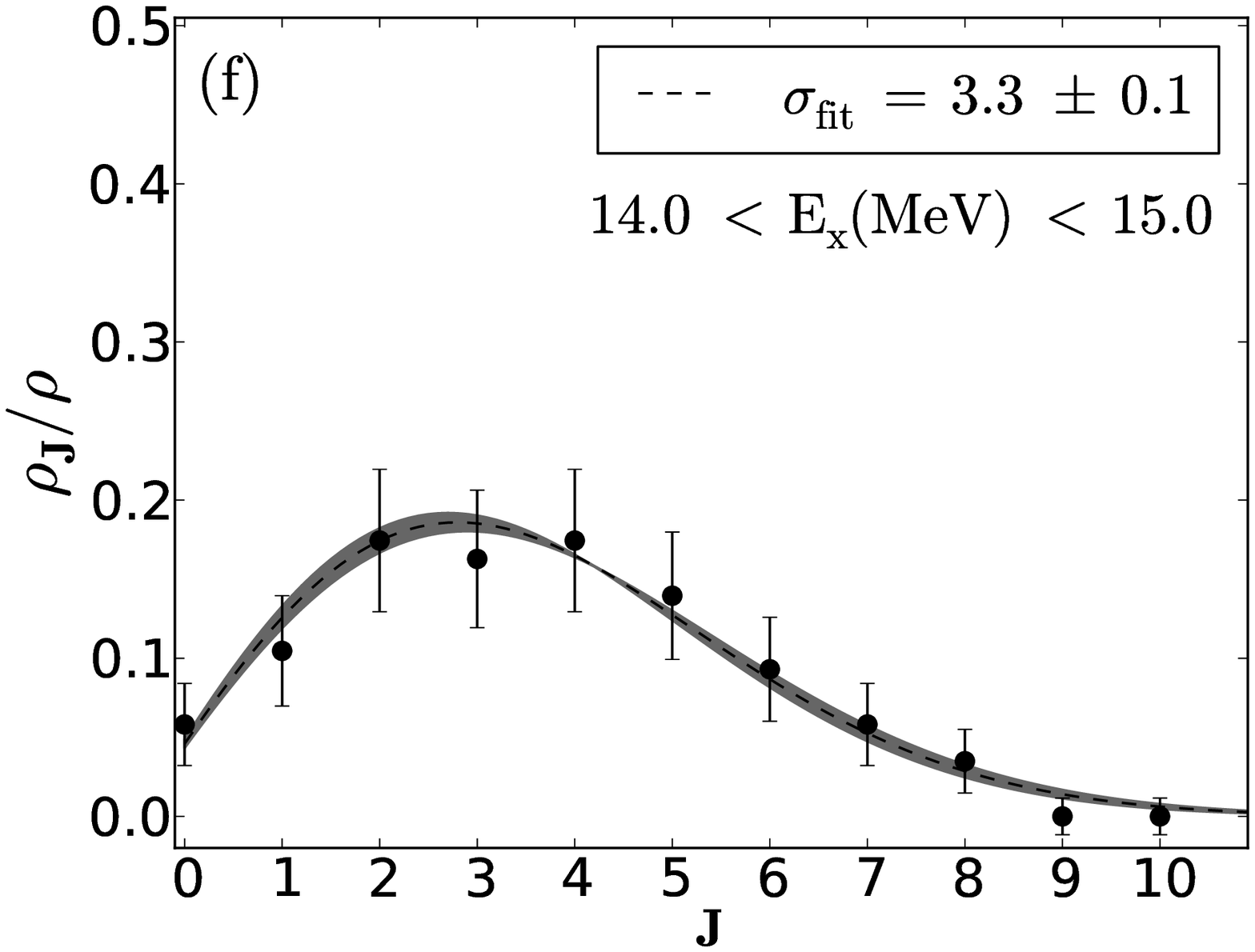}
}

\caption{As in Fig.~\ref{Al26} but for $^{26}$Mg in the $sd$ shell  \label{Mg26}}
\end{figure}

\begin{figure}
\subfloat{
\includegraphics[width=0.4\textwidth]{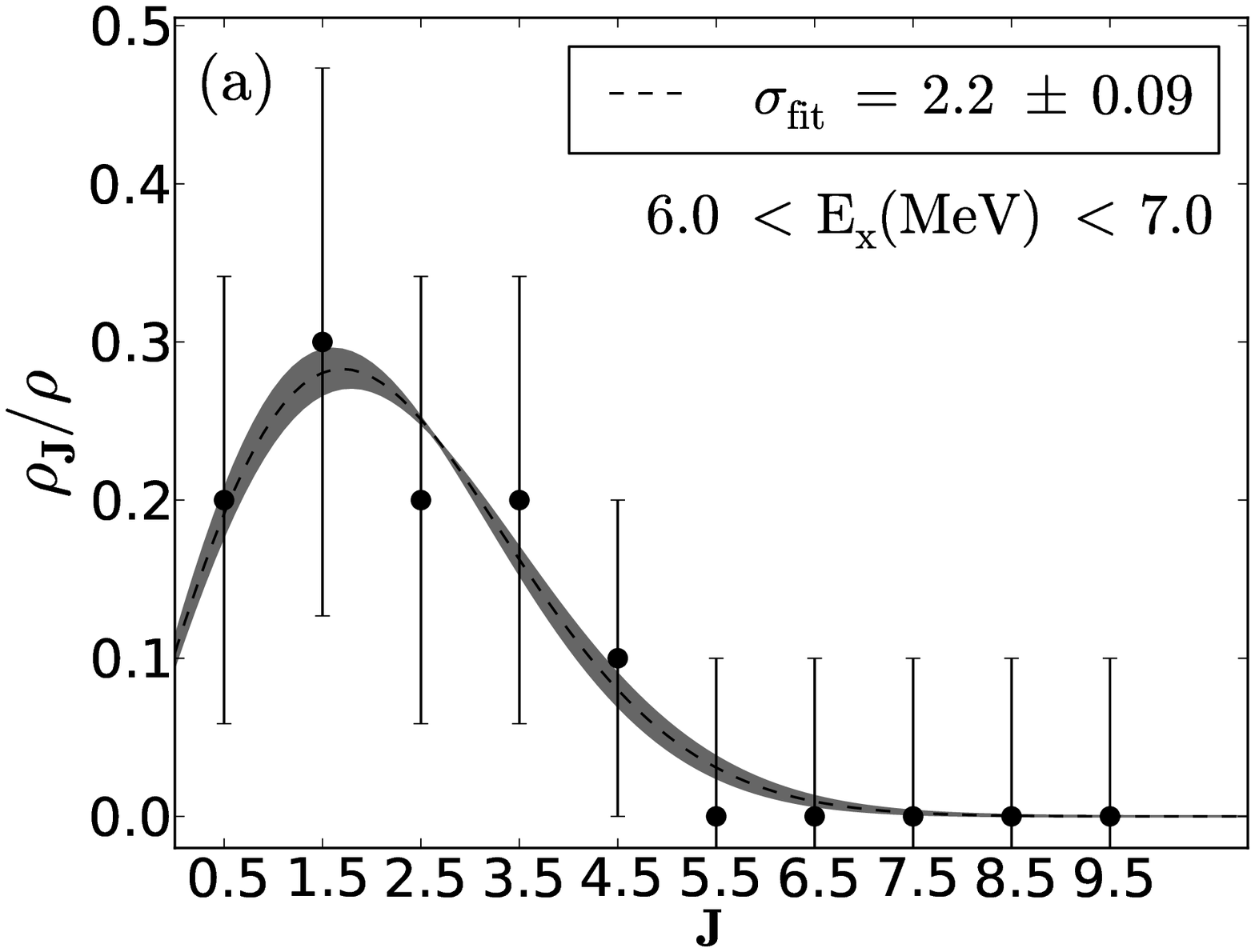}
}
\subfloat{
\includegraphics[width=0.4\textwidth]{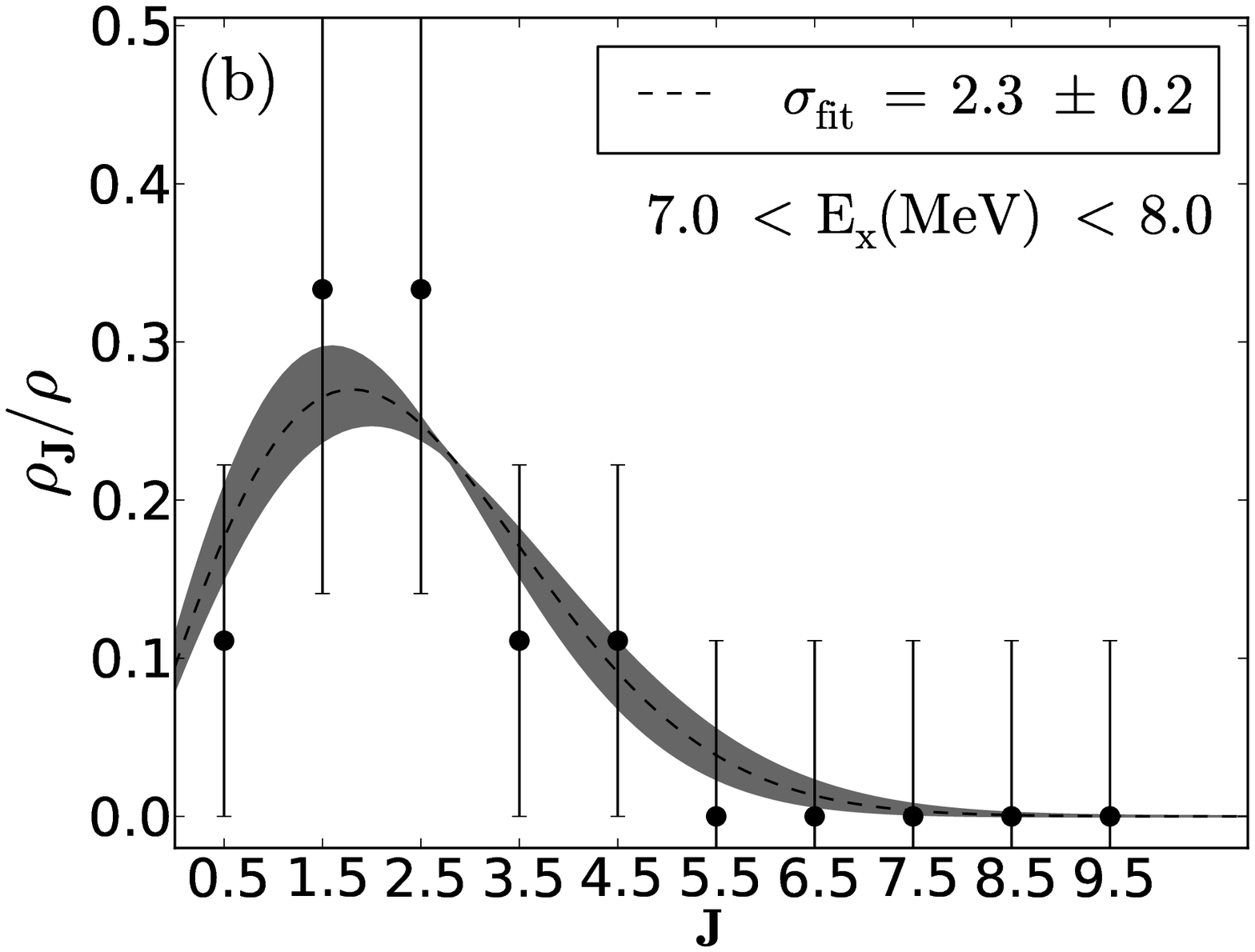}
} \\

\subfloat{
\includegraphics[width=0.4\textwidth]{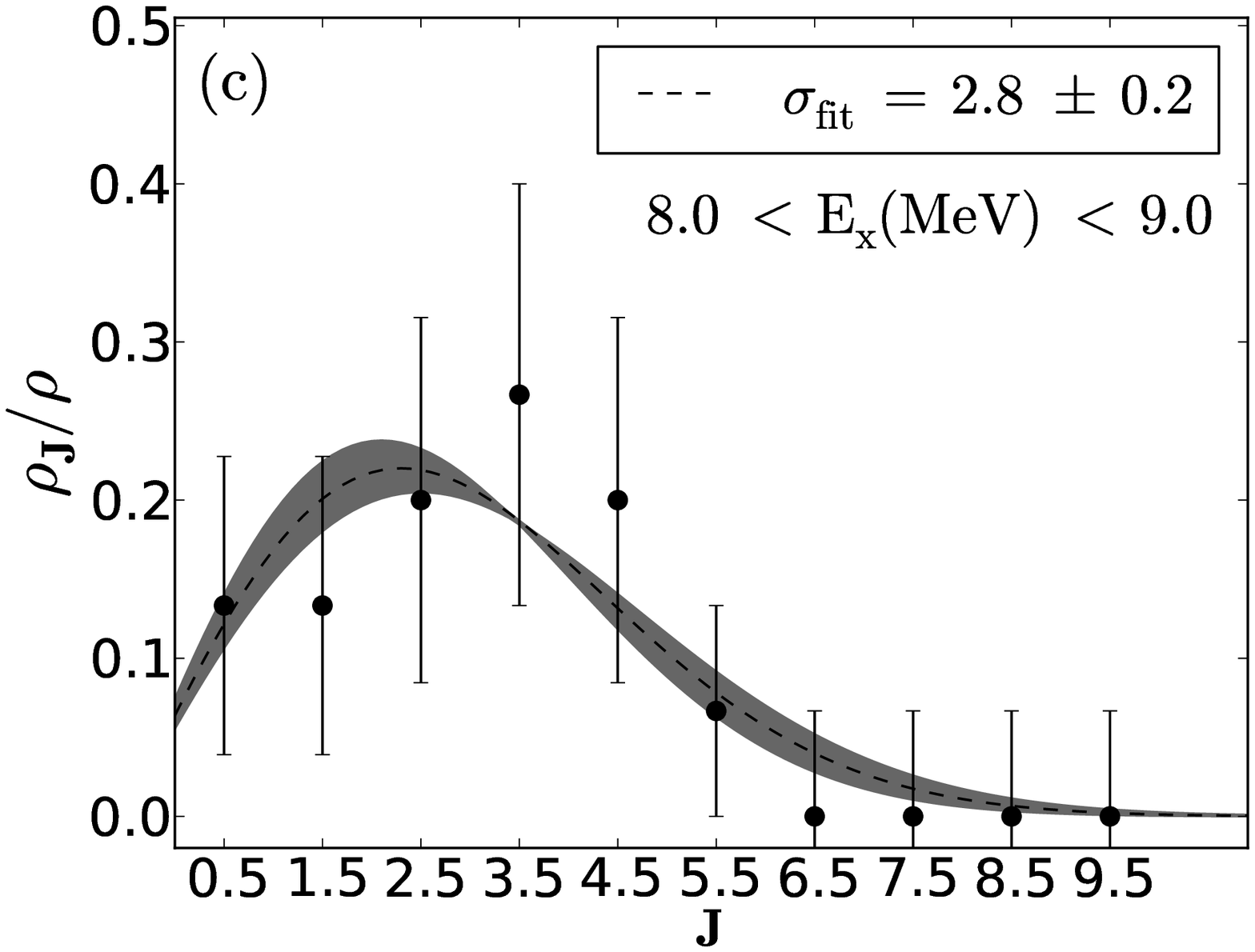}
}
\subfloat{
\includegraphics[width=0.4\textwidth]{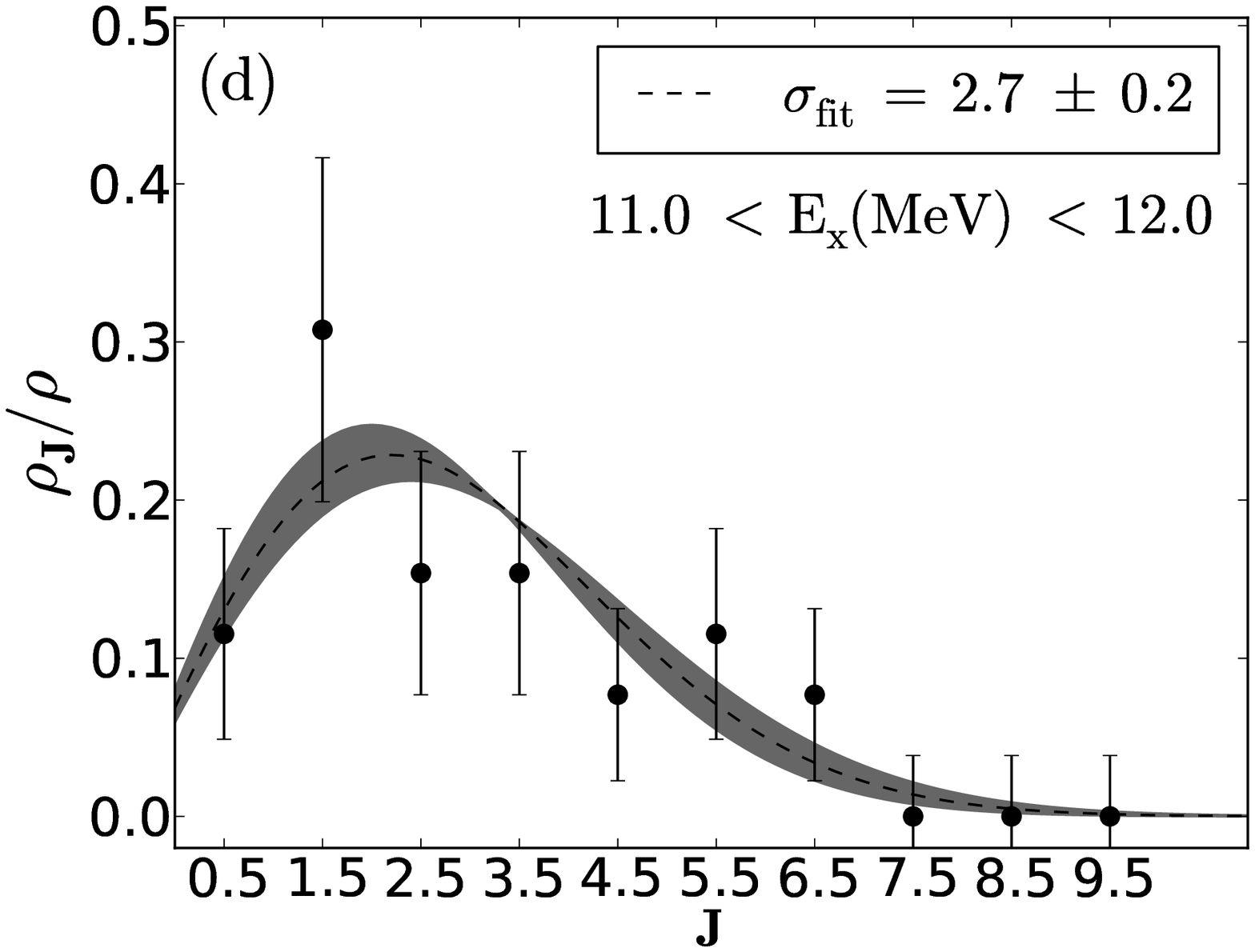}
} \\
\subfloat{
\includegraphics[width=0.4\textwidth]{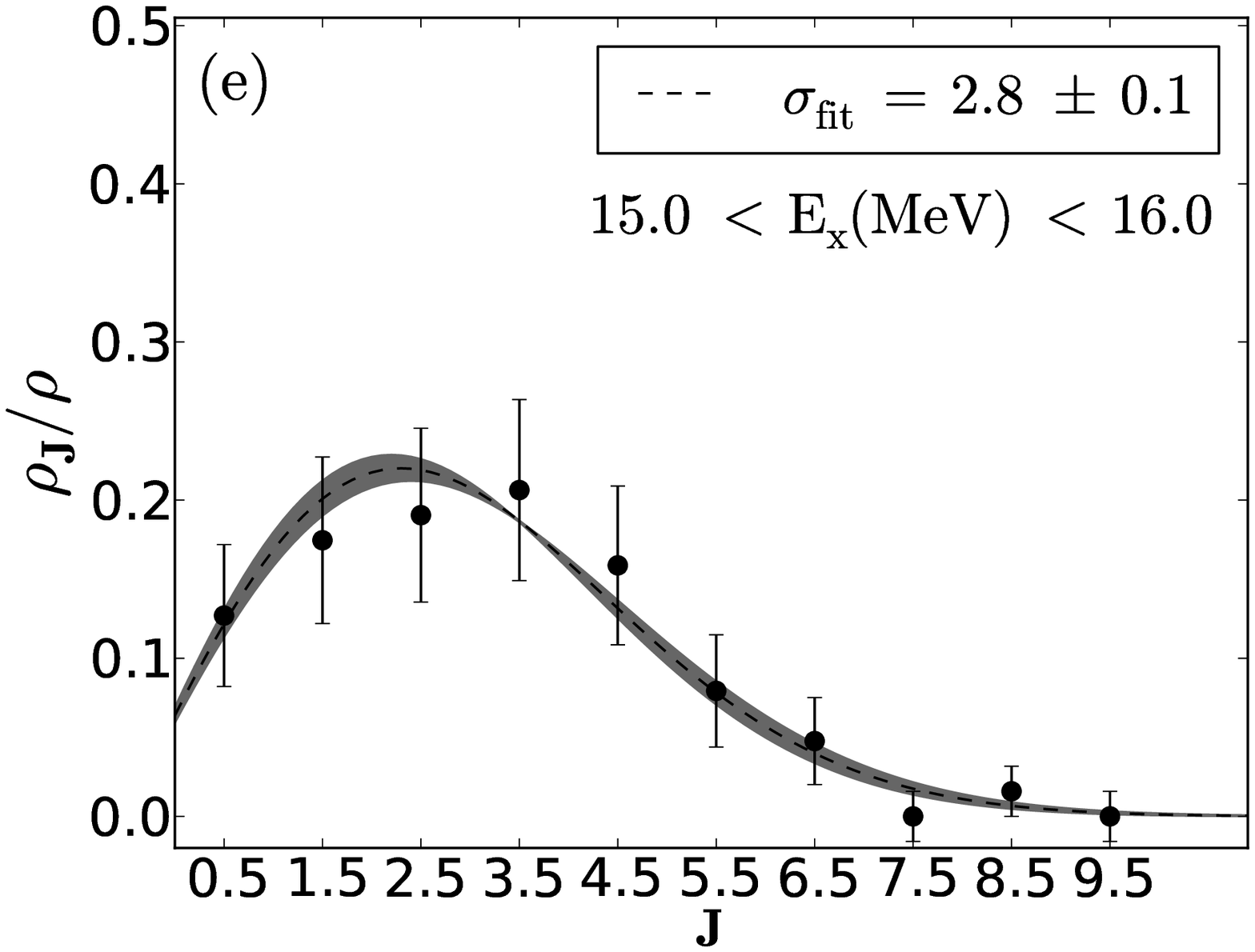}
}
\subfloat{
\includegraphics[width=0.4\textwidth]{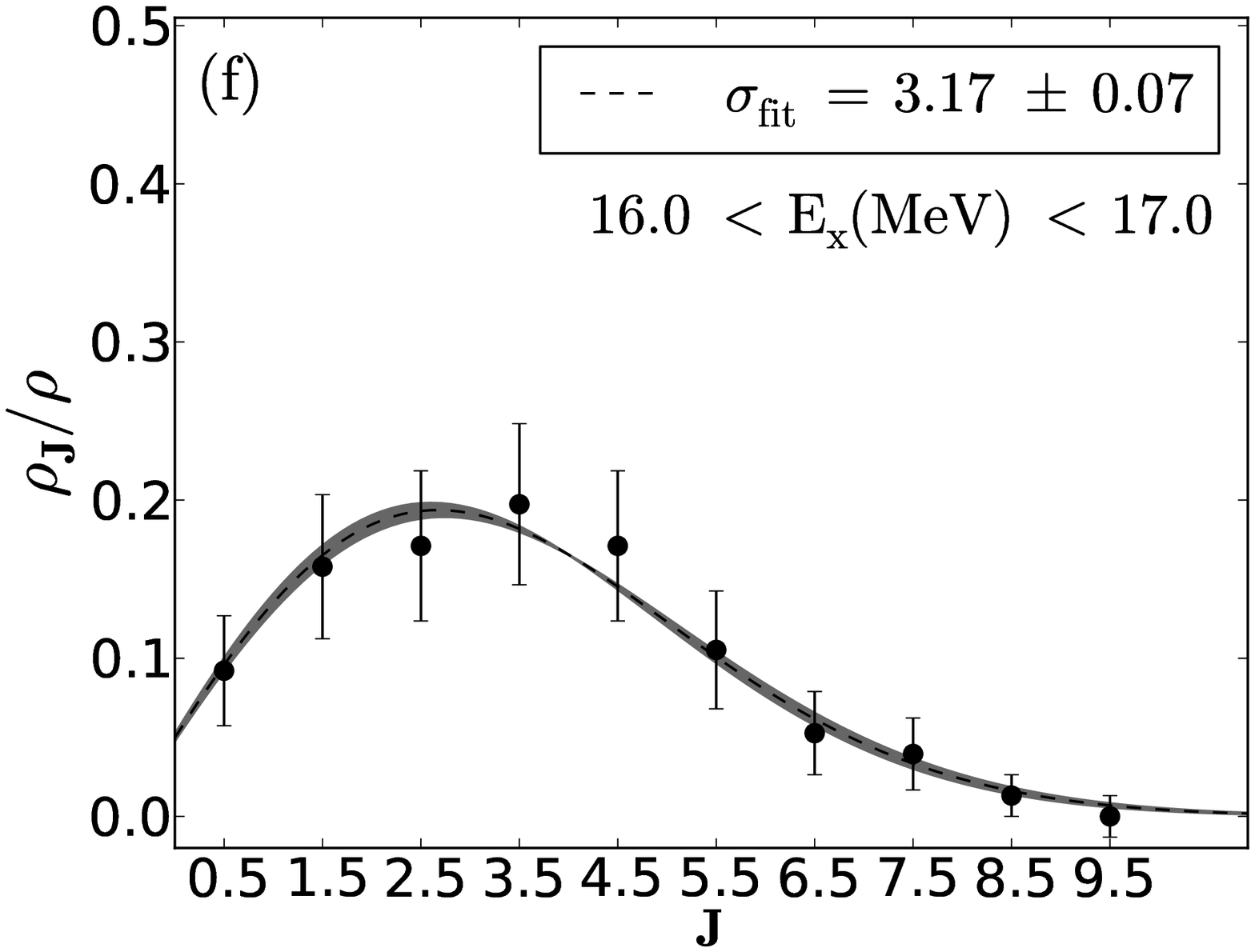}
}

\caption{As in Fig.~\ref{Al26} but for $^{33}$P in the $sd$ shell  \label{P33}}
\end{figure}

We begin in the $sd$-shell, where we consider 
$^{26}$Al, $^{26}$Mg, and $^{33}$P in Figs.~\ref{Al26}, \ref{Mg26}, and \ref{P33}, respectively, 
showing the distribution of states as a function of $J$.   Similar graphs for all nuclides listed in 
Table \ref{tab:list} can be found in \cite{SJsuppl}. 
Because we extract error bars on our spin-cutoff factors, we display not only the least-squares fit 
but also the one-standard-deviation envelope (the dark gray band in the figures). 
The points represent our binned CI shell-model data and the 
error bars  represent the statistical error $\sqrt{\rho_J}$ 
on the counts in a 1.0 MeV bin.  Not only do the error bars on the CI data get smaller, as the 
number of counts in the bins increase, our error on $\sigma$ decreases as well.

\begin{figure}

\subfloat{
\includegraphics[width=0.4\textwidth]{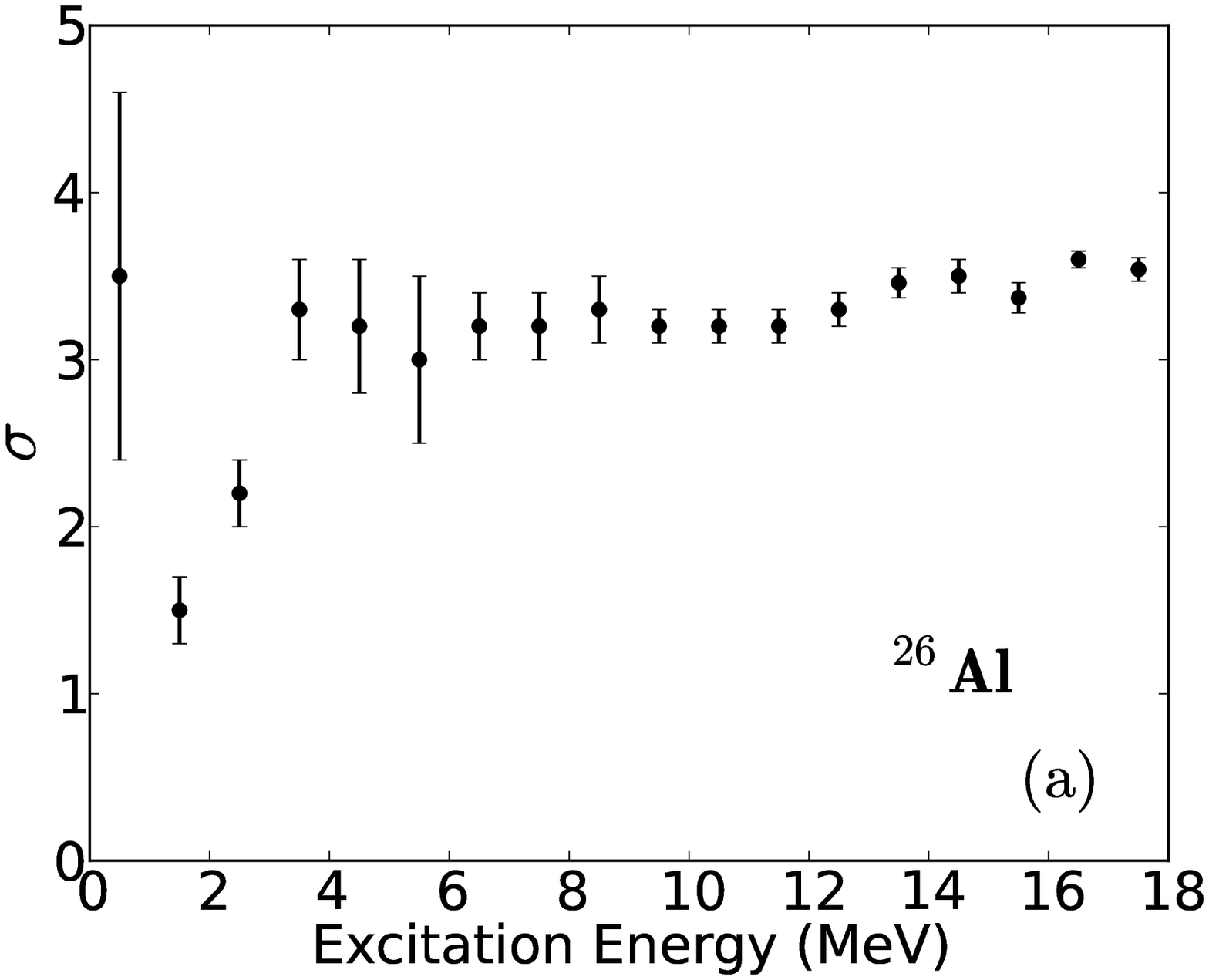}
}
\subfloat{
\includegraphics[width=0.4\textwidth]{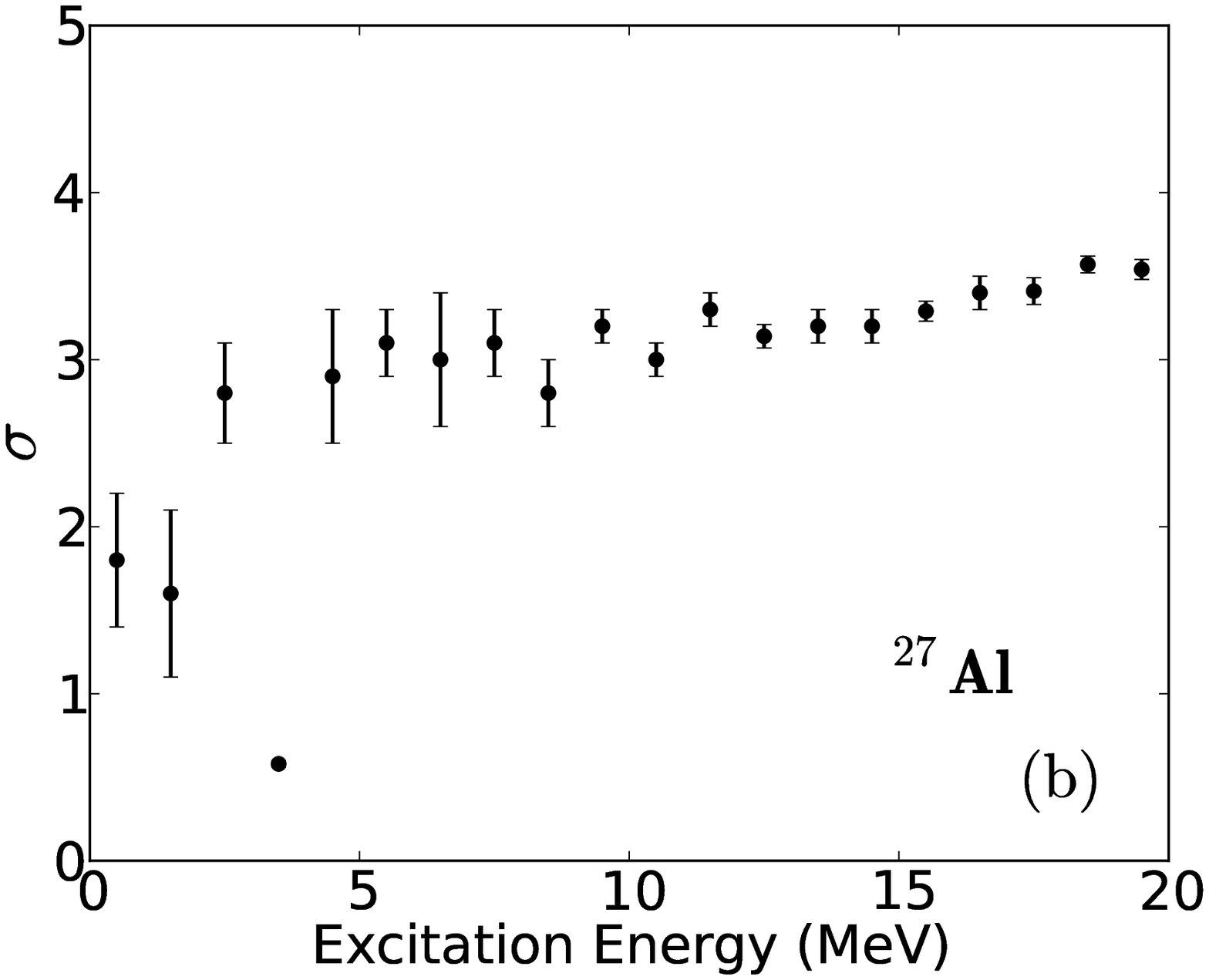}
} \\

\subfloat{
\includegraphics[width=0.4\textwidth]{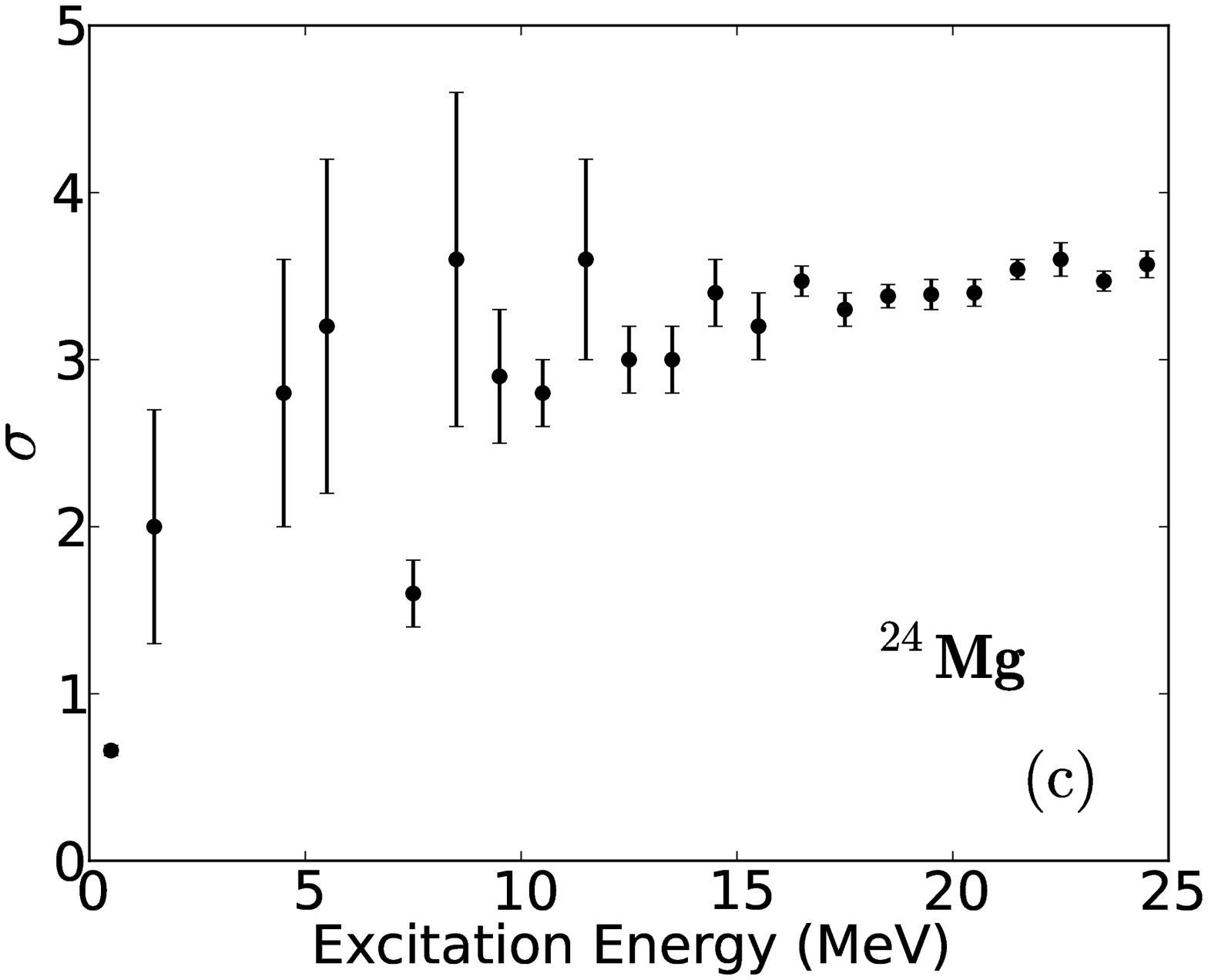}
}
\subfloat{
\includegraphics[width=0.4\textwidth]{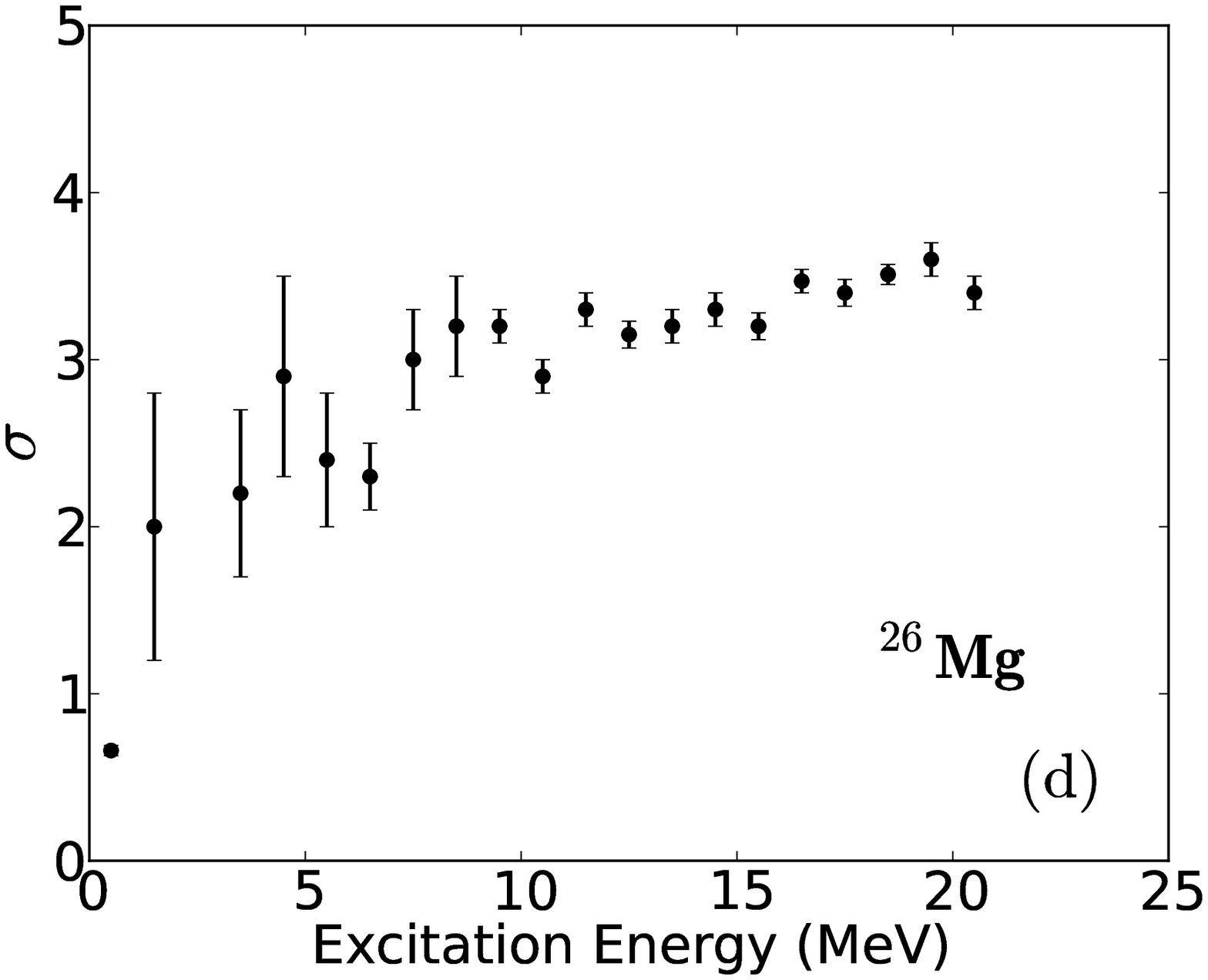}
} \\

\subfloat{
\includegraphics[width=0.4\textwidth]{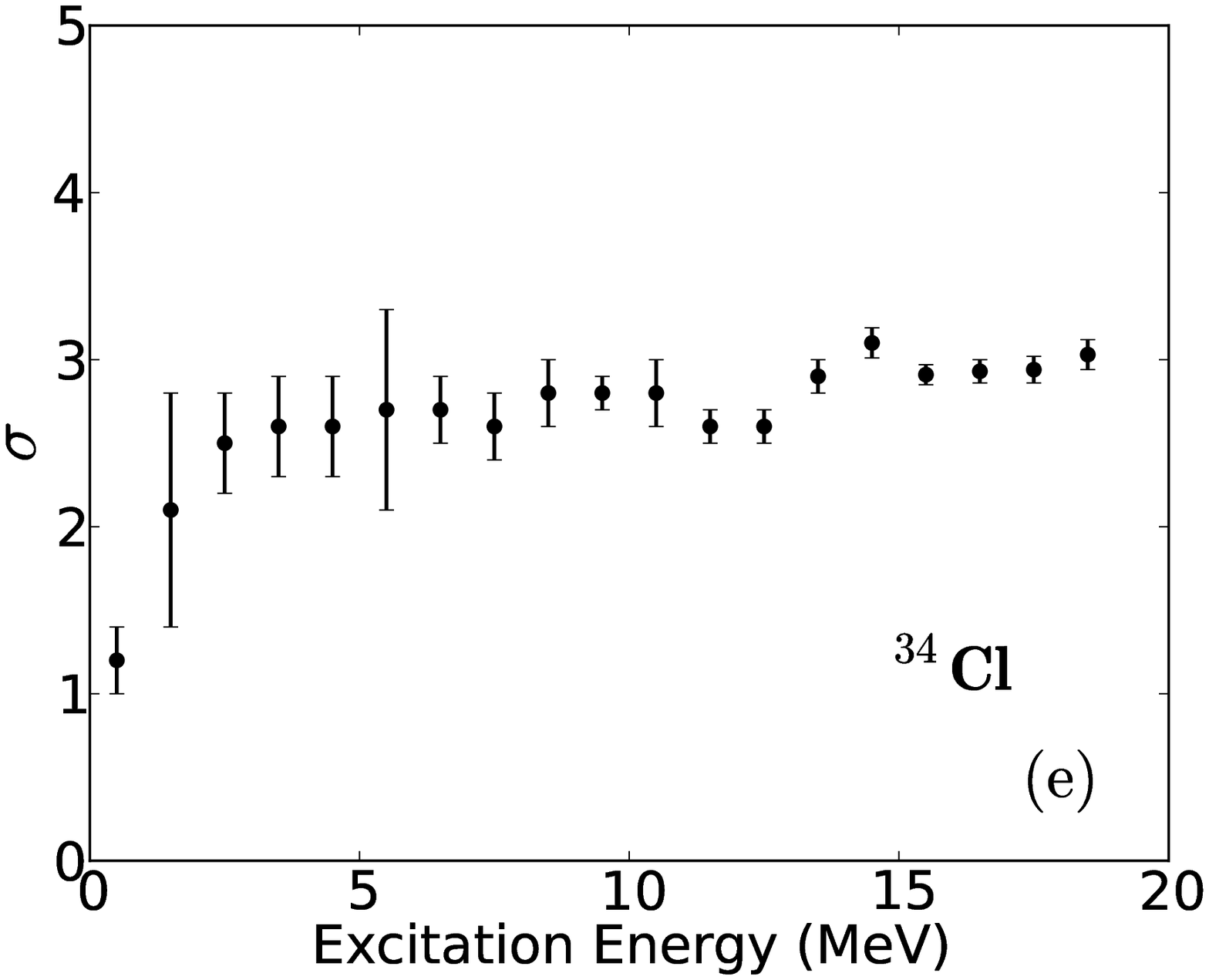}
}
\subfloat{
\includegraphics[width=0.4\textwidth]{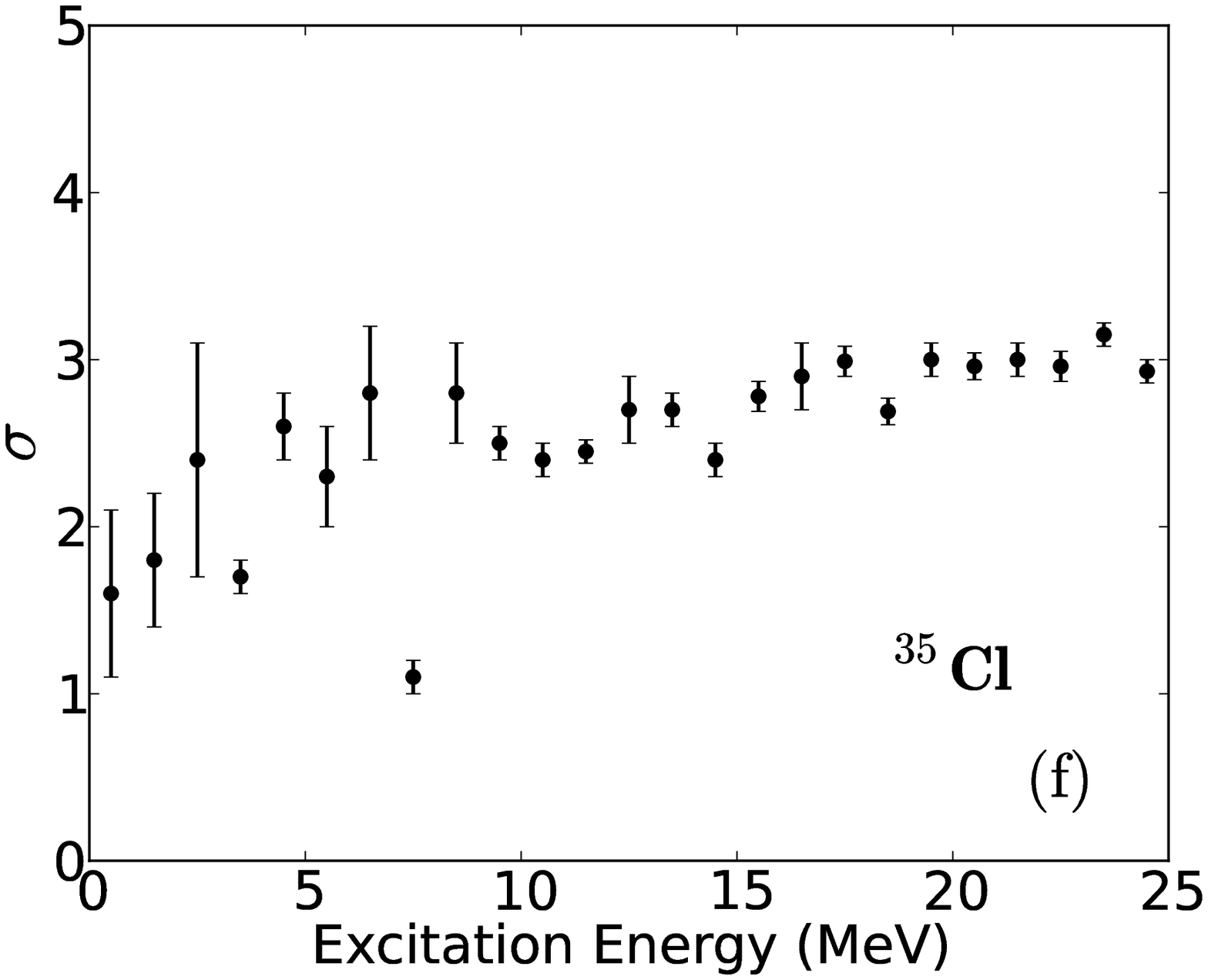}
} 
\caption{Evolution with excitation energy of fit spin-cutoff factors for selected $sd$-shell nuclides, 
using 1 MeV bins.}
\label{sd_scf}
\end{figure}

\begin{figure}

\subfloat{
\includegraphics[width=0.4\textwidth]{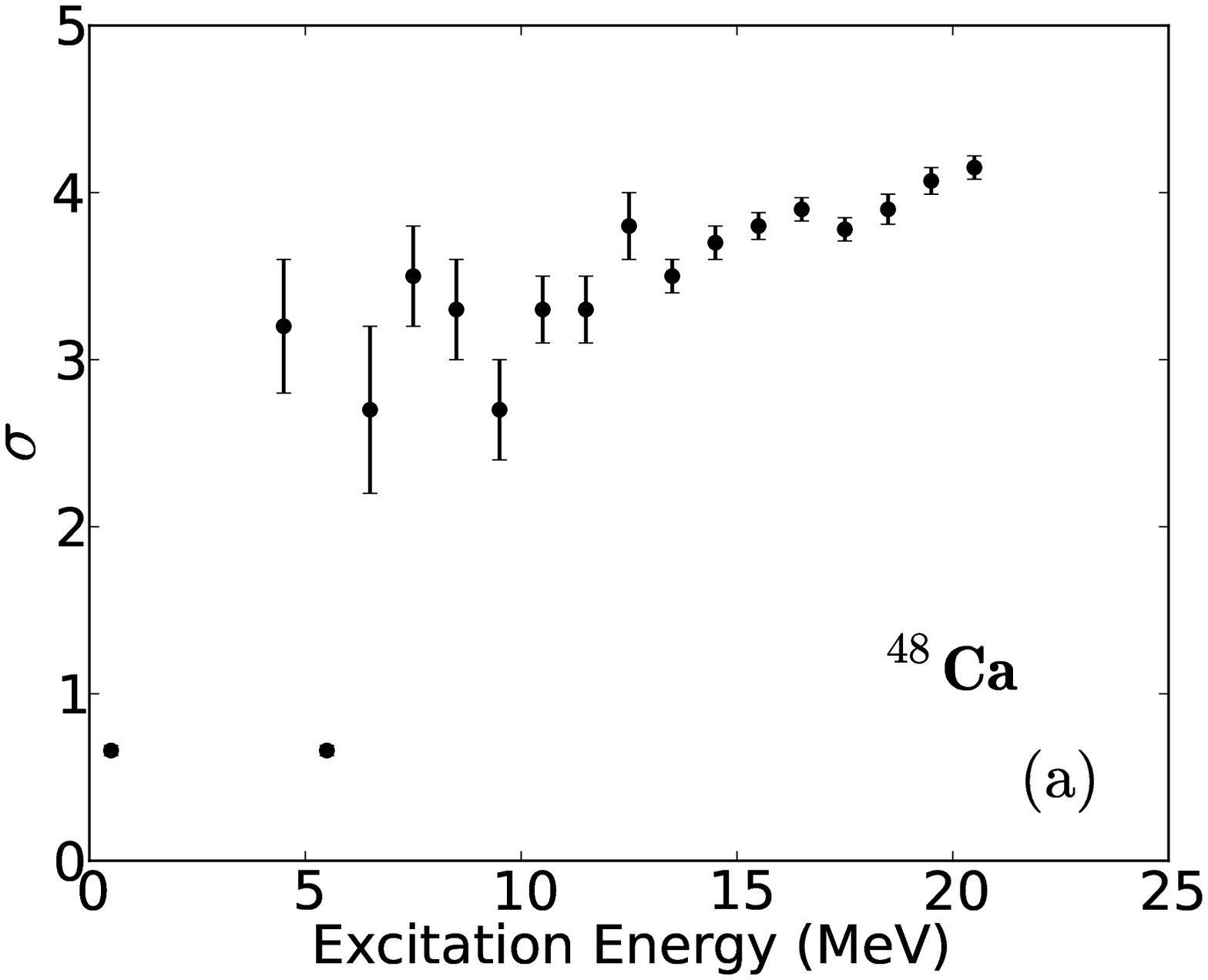}
}
\subfloat{
\includegraphics[width=0.4\textwidth]{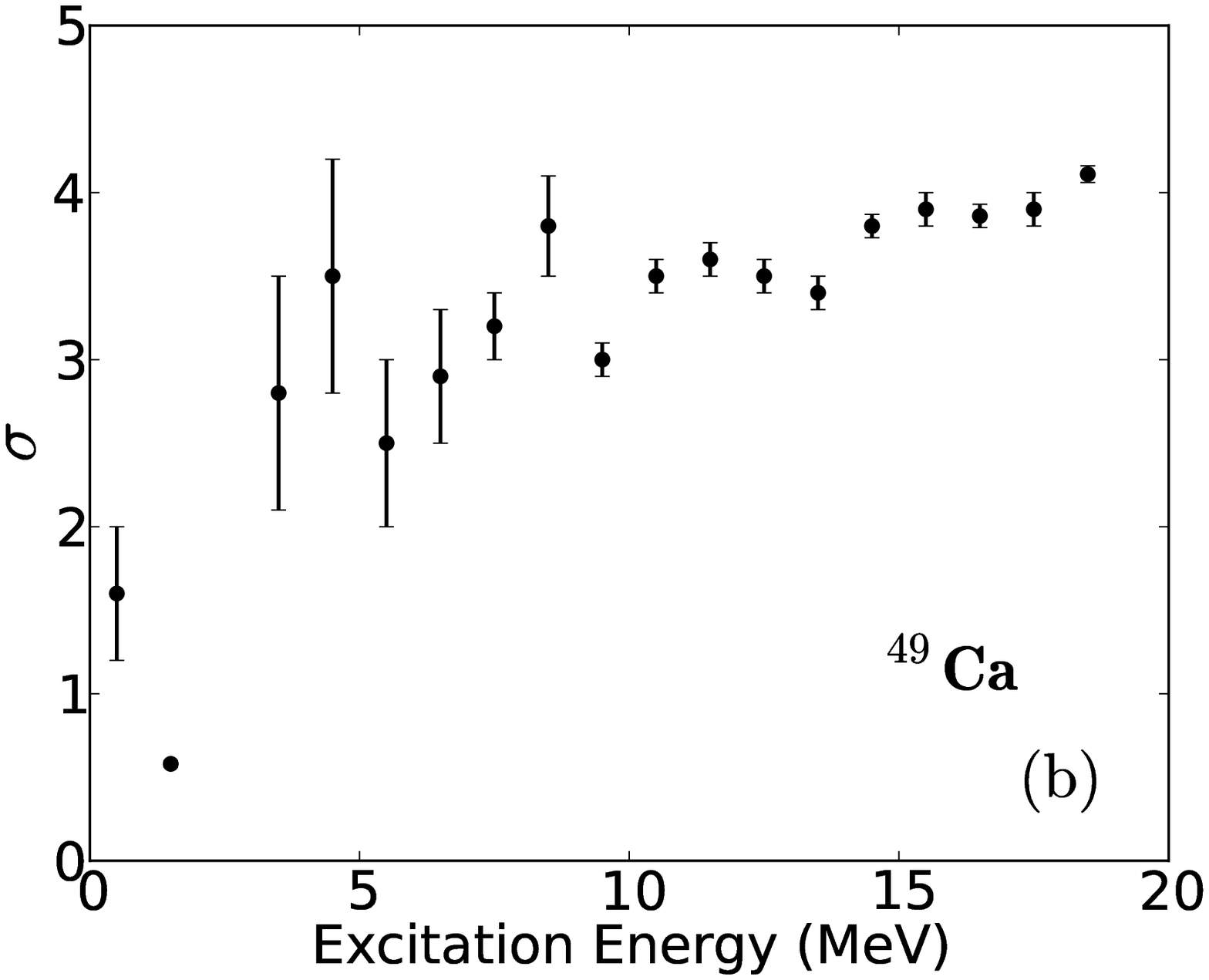}
} \\

\subfloat{
\includegraphics[width=0.4\textwidth]{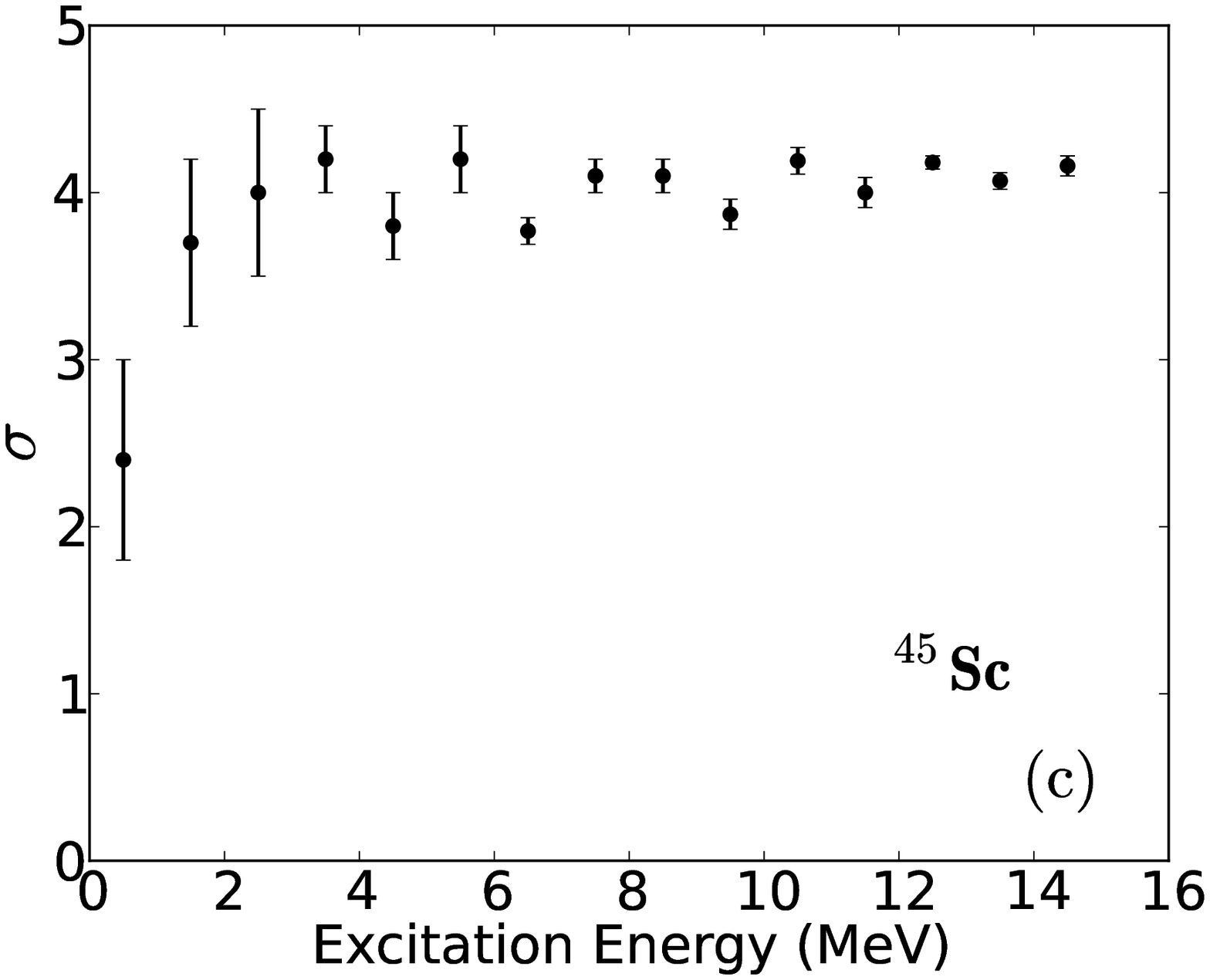}
}
\subfloat{
\includegraphics[width=0.4\textwidth]{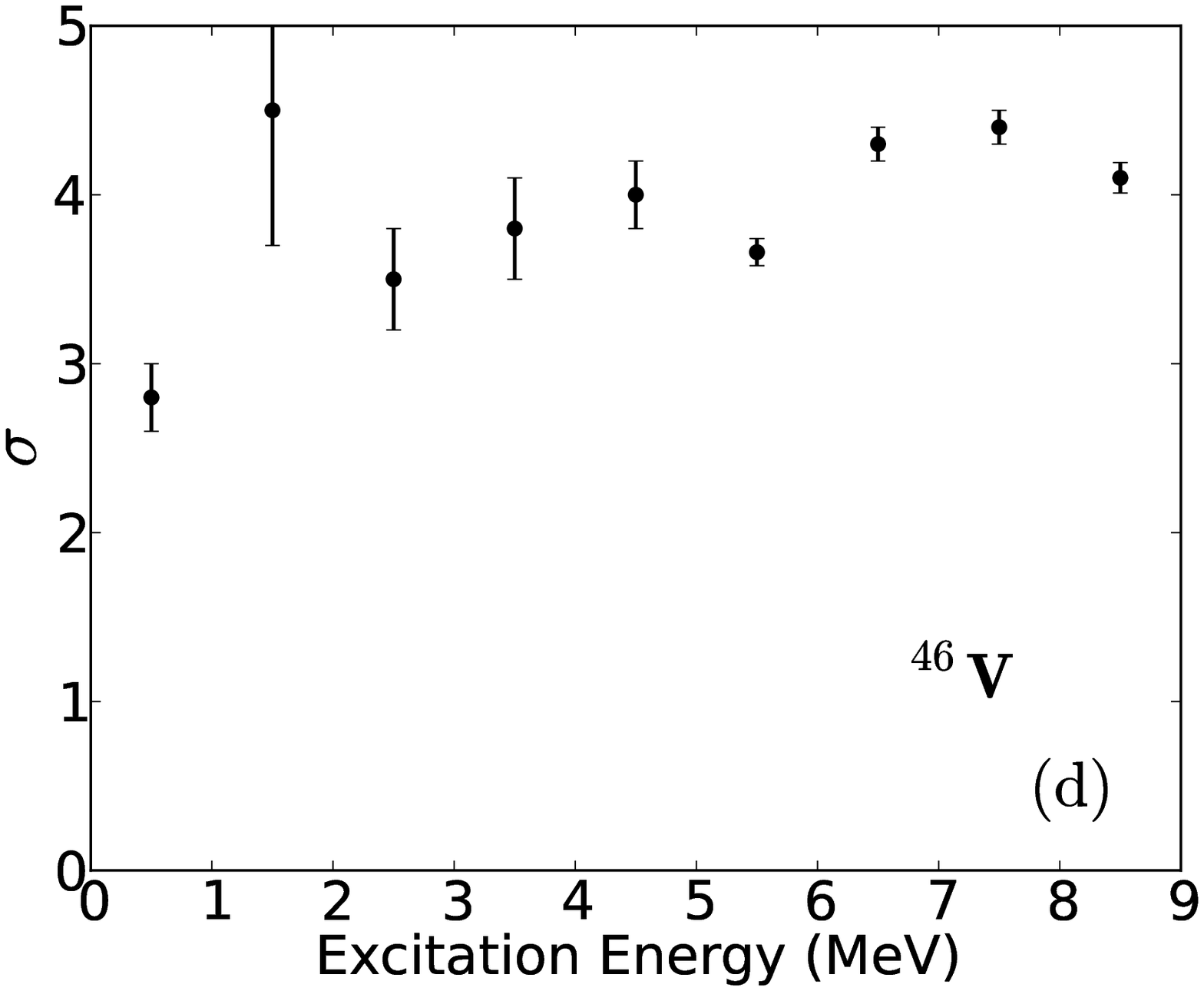}
} \\

\subfloat{
\includegraphics[width=0.4\textwidth]{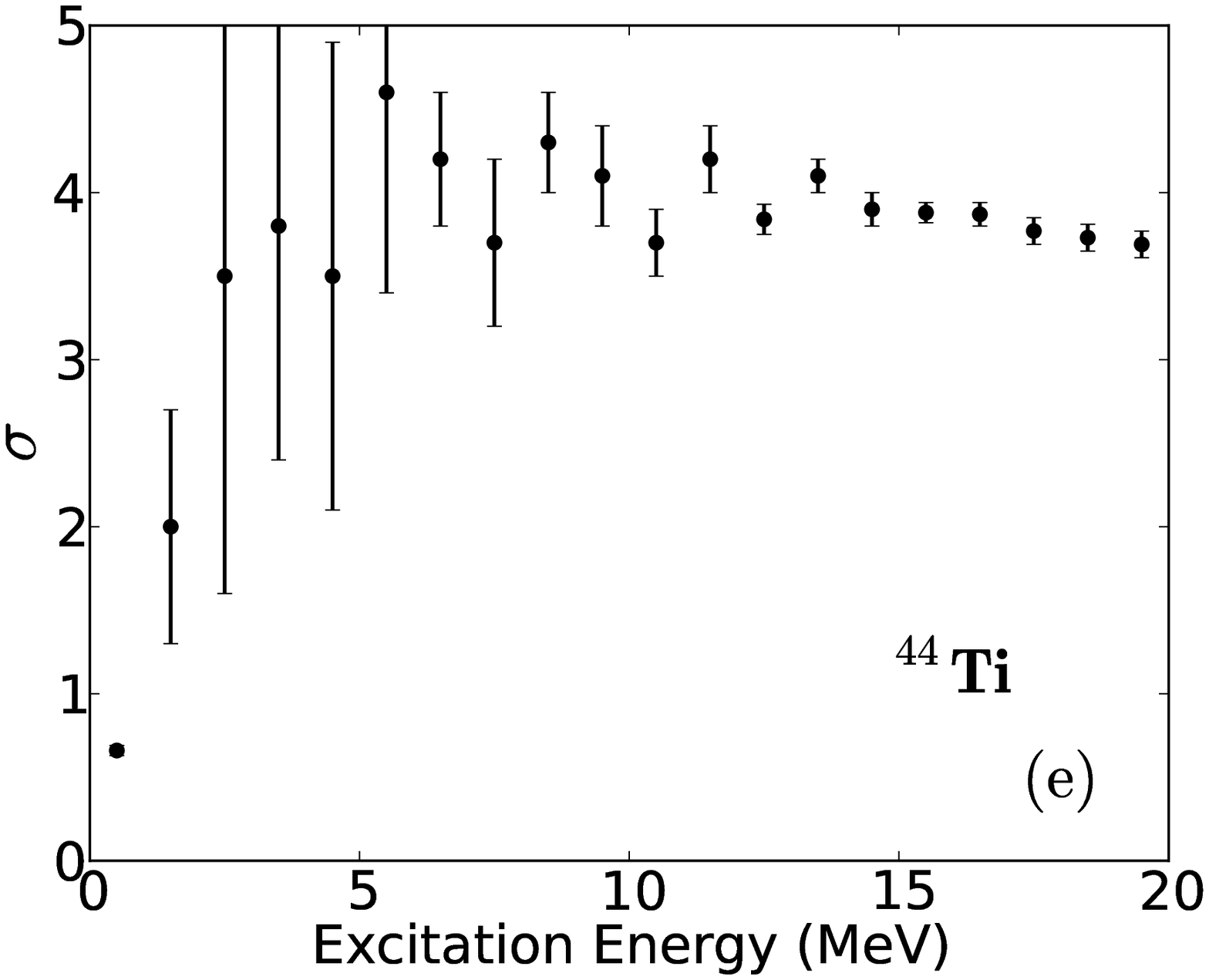}
}
\subfloat{
\includegraphics[width=0.4\textwidth]{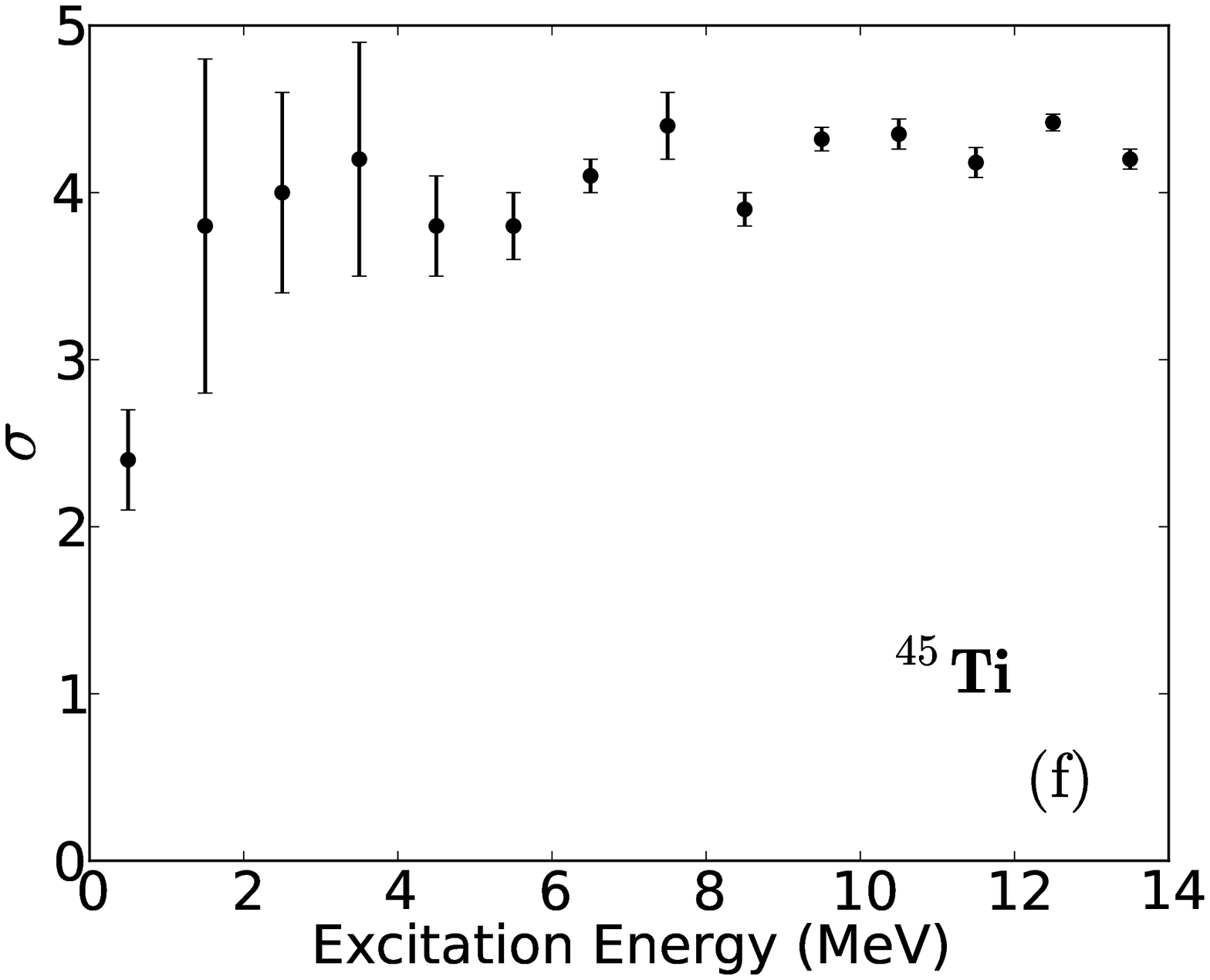}
} 
\caption{Evolution with excitation energy of fit spin-cutoff factors for selected $pf$-shell nuclides, 
using 1 MeV bins.}
\label{pf_scf}
\end{figure}

\begin{figure}

\subfloat{
\includegraphics[width=0.4\textwidth]{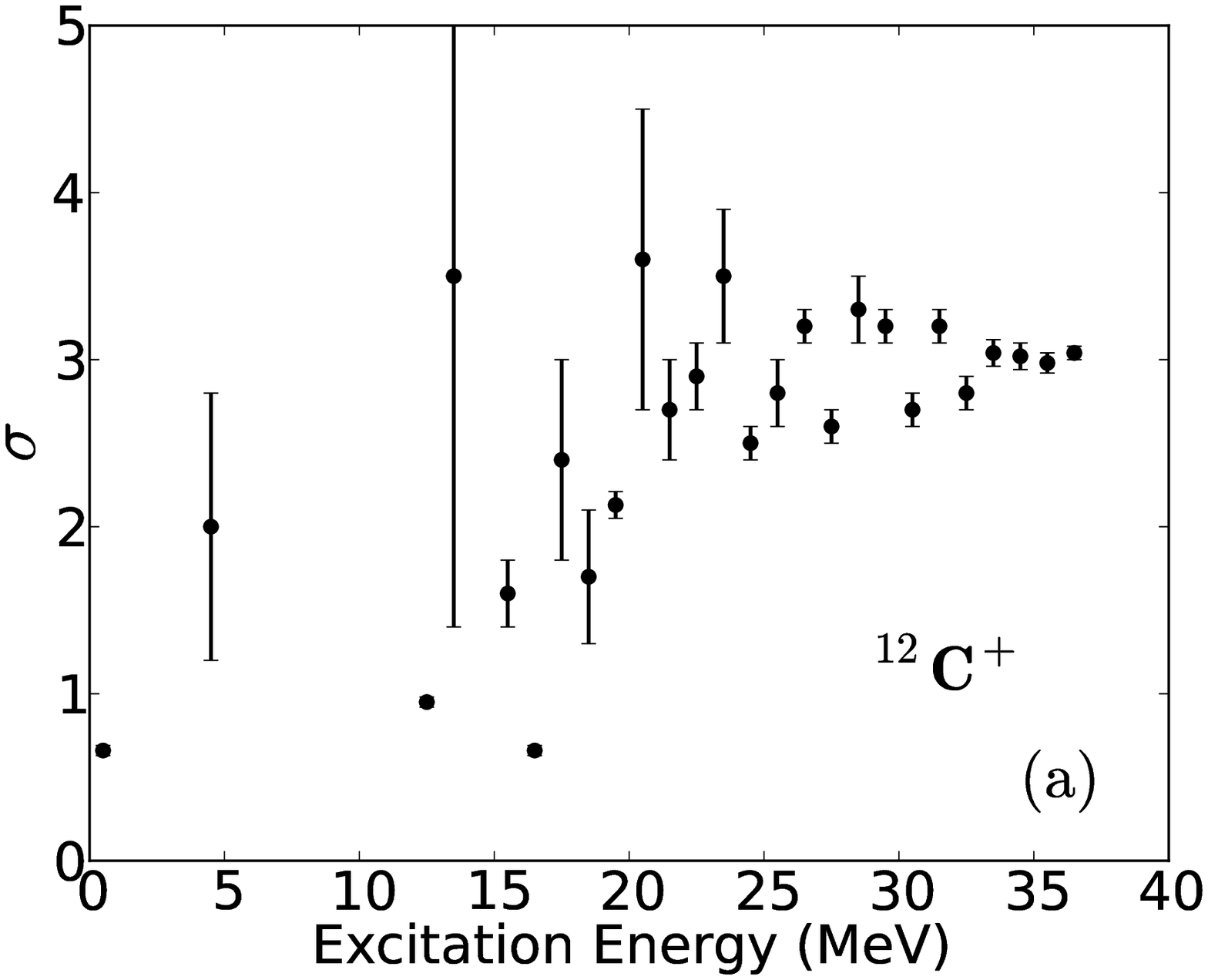}
}
\subfloat{
\includegraphics[width=0.4\textwidth]{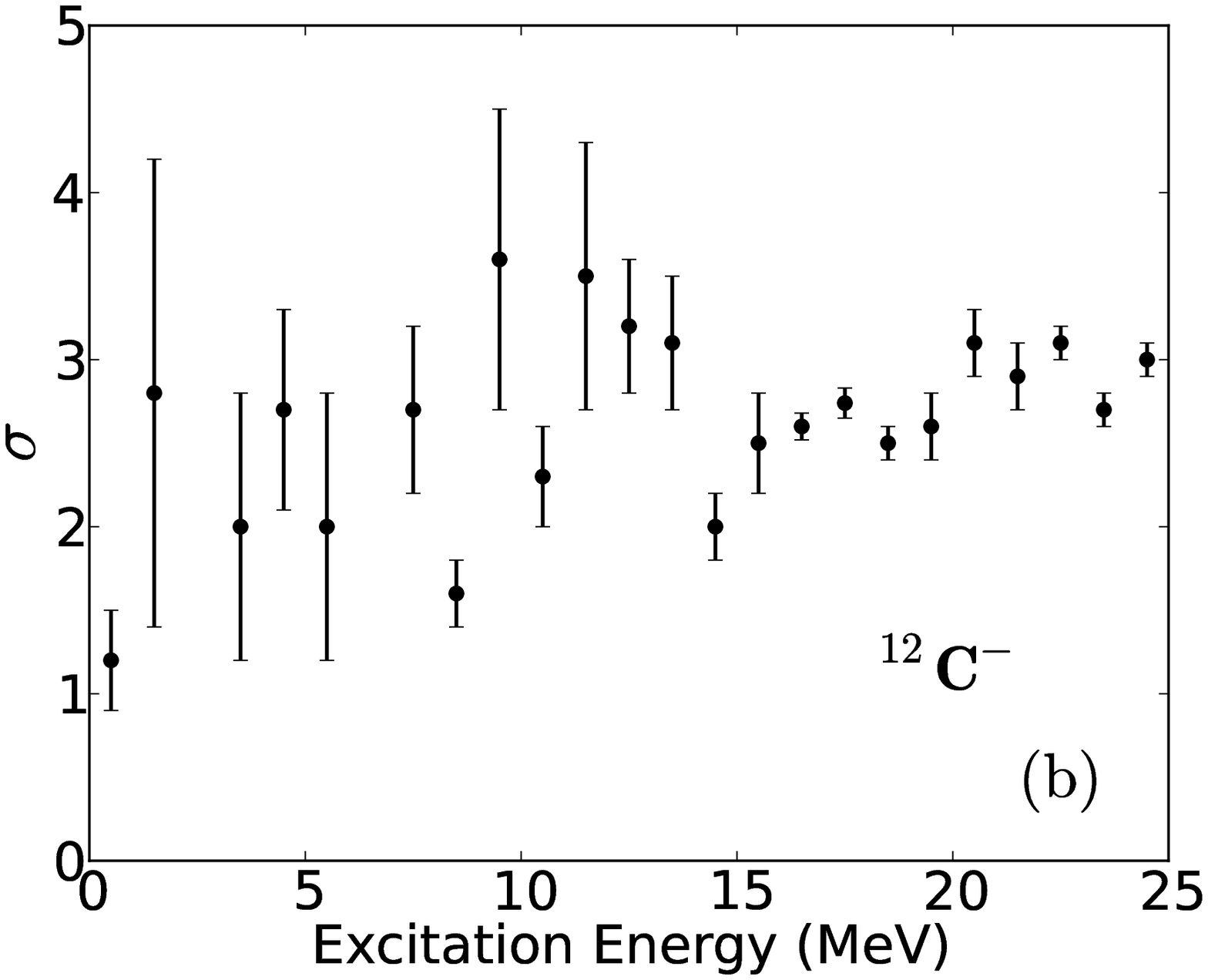}
} \\

\subfloat{
\includegraphics[width=0.4\textwidth]{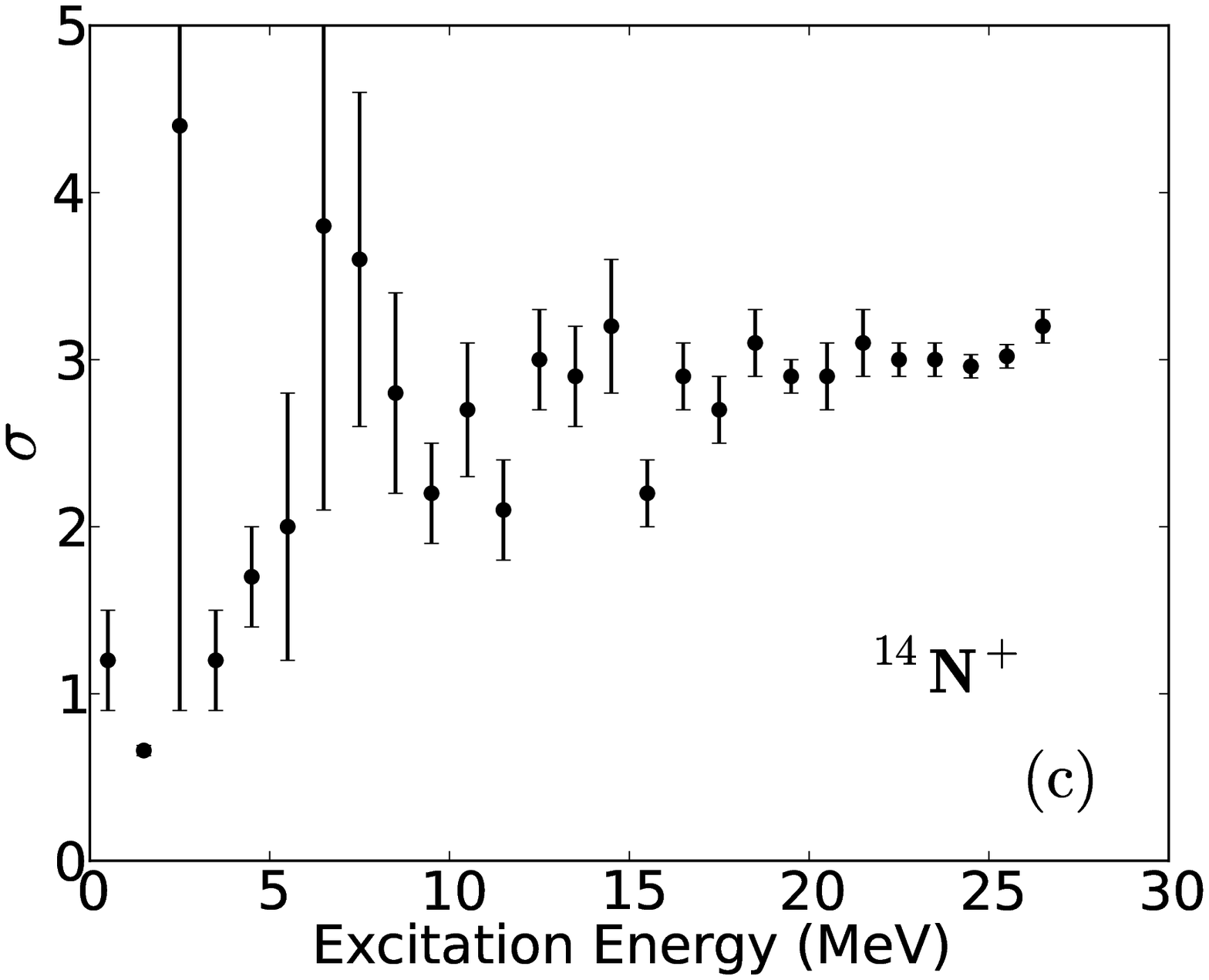}
}
\subfloat{
\includegraphics[width=0.4\textwidth]{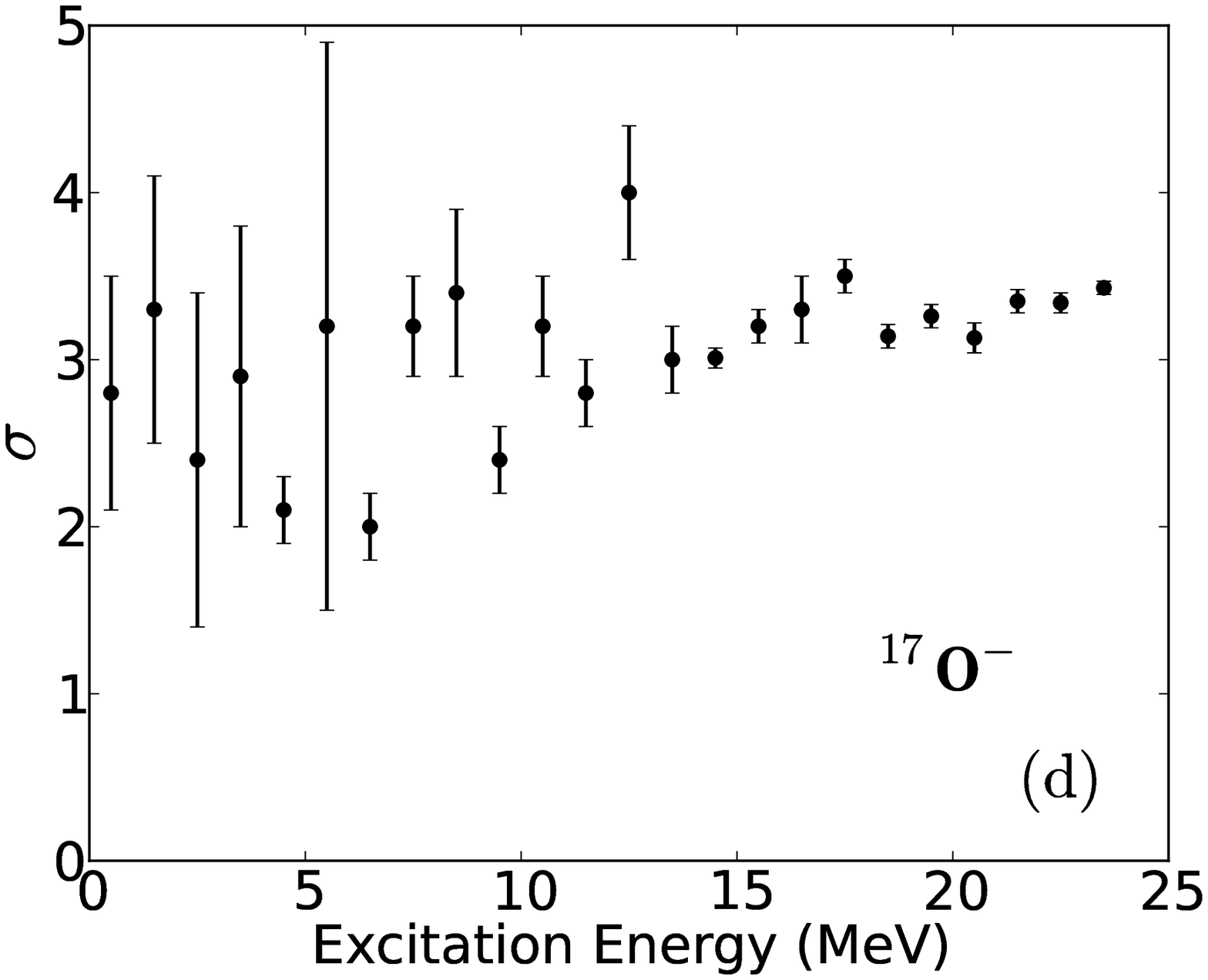}
} \\

\subfloat{
\includegraphics[width=0.4\textwidth]{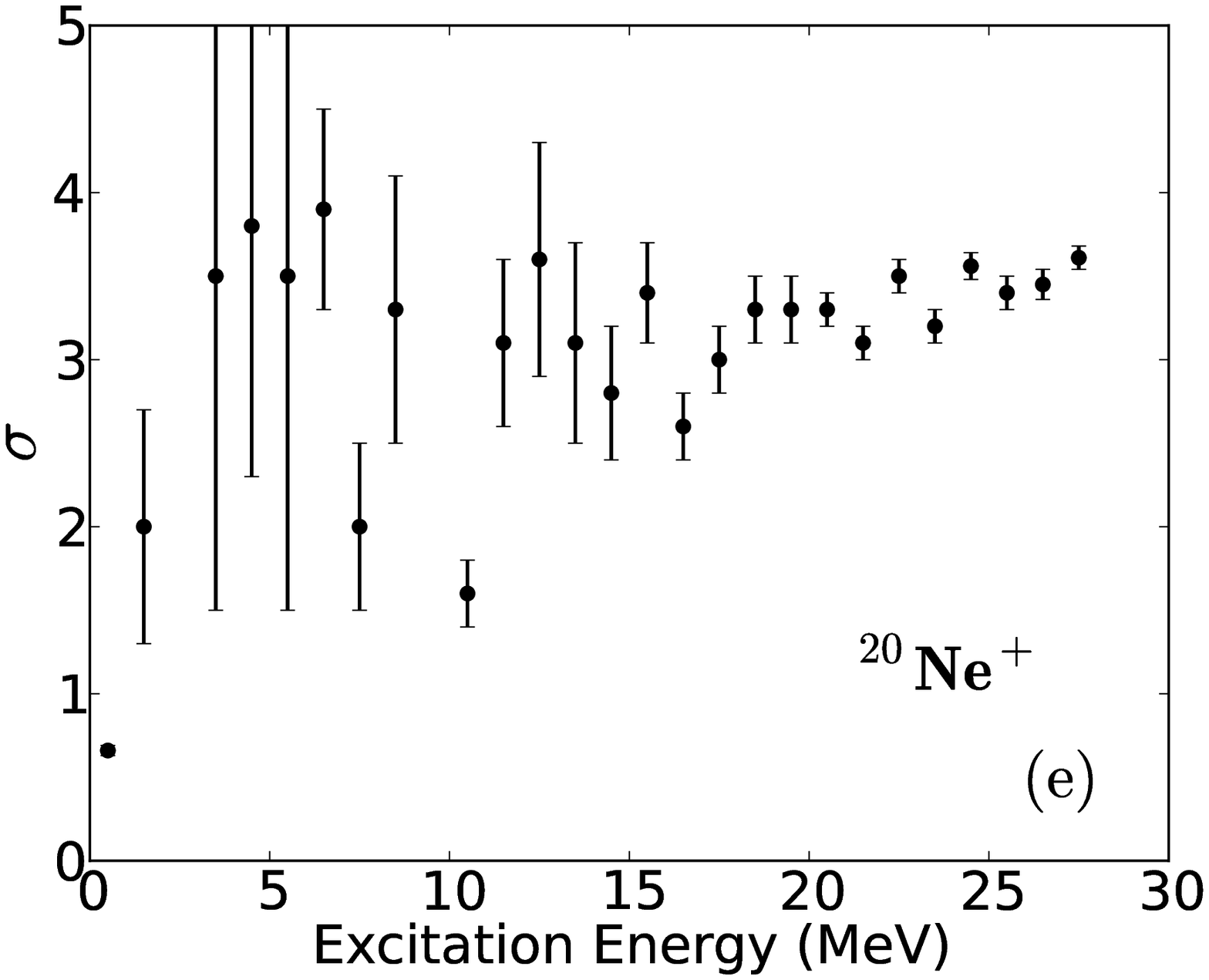}
}
\subfloat{
\includegraphics[width=0.4\textwidth]{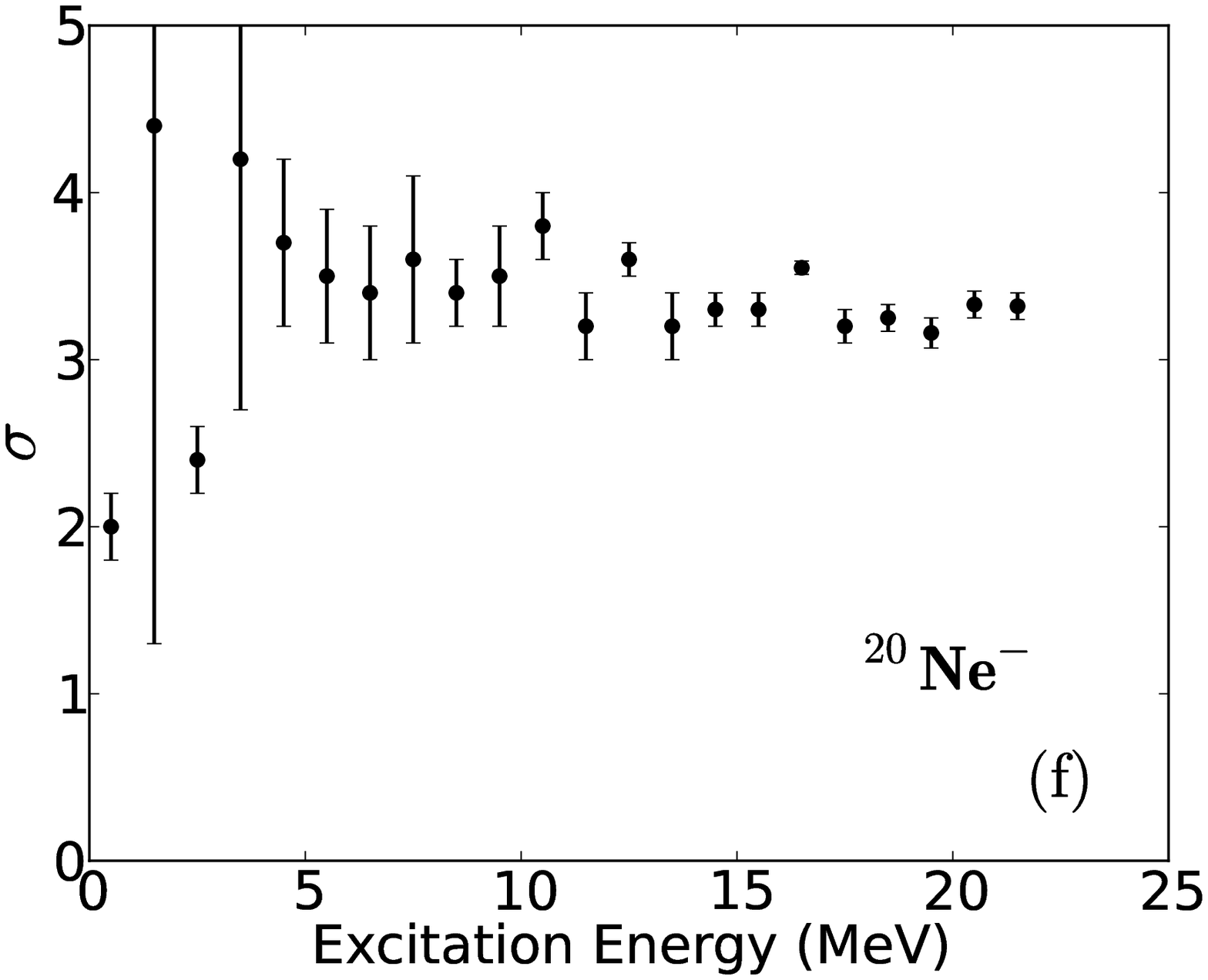}
} 
\caption{Evolution with excitation energy of fit spin-cutoff factors 
for selected $p$-$sd_{5/2}$-shell nuclides, using 1 MeV bins,
for both positive parity (left-hand figures) and negative parity (right-hand figures) }
\label{psd_scf}
\end{figure}

In Fig.~\ref{sd_scf} we show the evolution of the spin-cutoff factor as a function of excitation energy
for these nuclides as well 
as several additional $sd$-shell nuclides. Fig.~\ref{pf_scf} shows the same for $pf$ shell nuclides, 
while Fig.~\ref{psd_scf} show the evolution for both positive and negative parities in the $p$-$sd_{5/2}$ 
space. 
Additional graphs encompassing all nuclides listed in 
Table \ref{tab:list} can be found in \cite{SJsuppl}. 
The error bars are larger for the last case because the statistics are smaller, and the 
values of the level density in all cases are no doubt an underestimate because intruders are left out. 
We remind the reader our goal is not to determine the `true' or experimental value of the spin-cutoff factor but 
to compare the statistical goodness derived from semi-realistic calculations.

%\begin{figure}[H]
%	\centering
%	\includegraphics[width=0.5\textwidth]{Al_11.eps}
%	\caption{$^{26}$Al in the $sd$ shell,  for the energy bin between 11.0 and 12.0 MeV excitation energy.}
%\end{figure}

%\begin{figure}[ht] \centering

Figures \ref{sd_scf}, \ref{pf_scf}, and \ref{psd_scf} show that as excitation energy increases the
distribution in angular momentum is better described by the spin cut-off
parameterization, Eq.~(\ref{Ericson}), consequently resulting in the reduction of the
error on the extracted spin-cutoff factor.
The pattern of improvement in the fit is observed in all analyzed
 nuclides, and can be seen by examining the root mean
square error (RMSE) between the CI data and the fit:
\begin{equation} \label{eq:rmse}
	\left(\frac{\rho_{J}}{\rho}\right)_{\mathrm{RMSE}}
	= \left<\left[\left(\frac{\rho_{J}}{\rho}\right)_{\mathrm{shell}}
	-\left(\frac{\rho_{J}}{\rho}\right)_{\mathrm{fit}}\right]^{2}
	\right>_{J}^{1/2}.
\label{RMSE}
\end{equation}

\begin{figure}[ht]

\centering
		\includegraphics[width=0.7\textwidth]{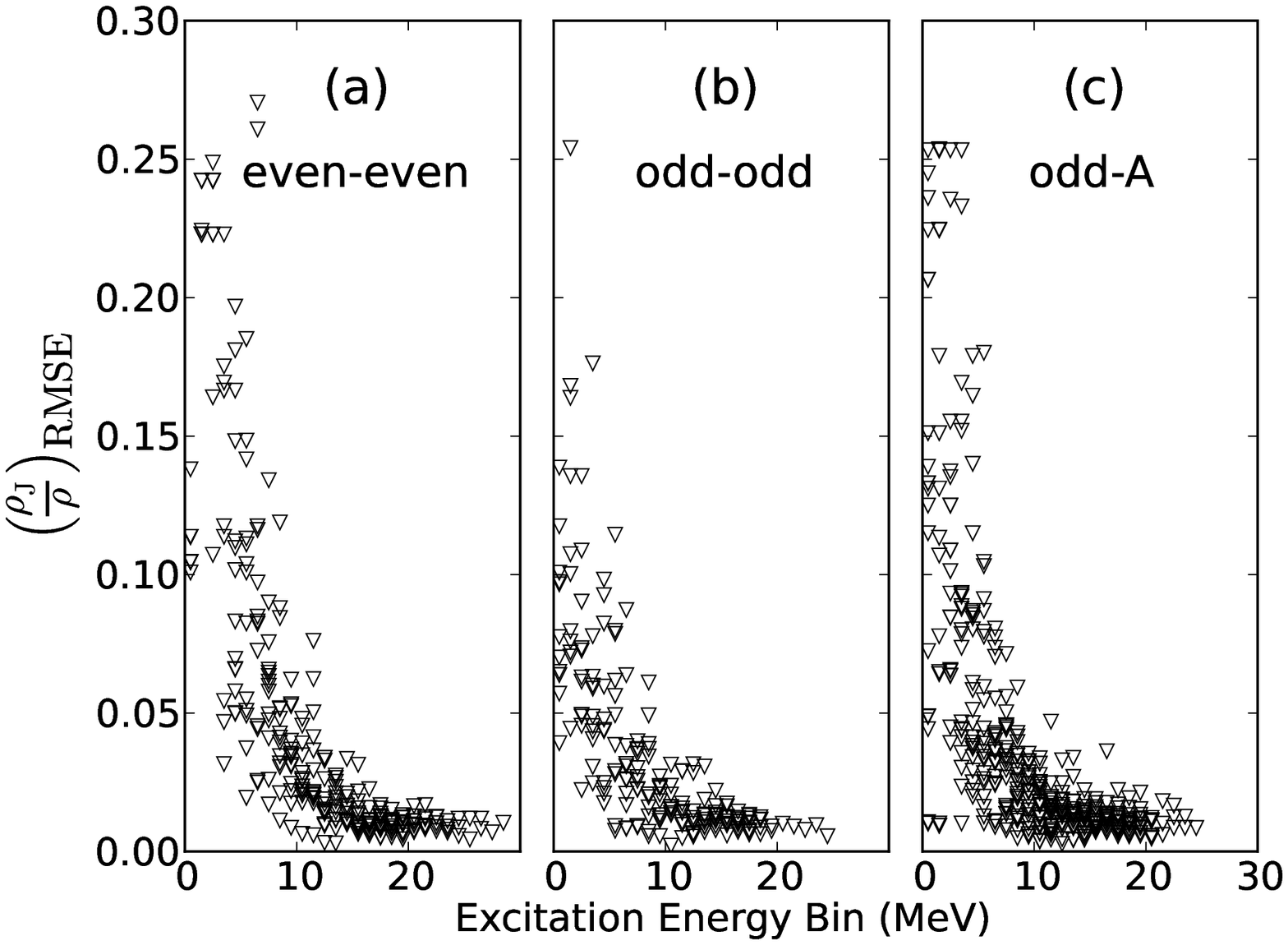}

	\caption{$(\rho_{J}/\rho)_{\mathrm{RMSE}}$ (defined in Eq.~\ref{RMSE})  for all analyzed
\textit{sd} and \textit{pf} shell nuclides organized by structure.\label{fig:err_struc}}
\end{figure}

\begin{figure}[ht]

\centering

		\includegraphics[width=0.7\textwidth]{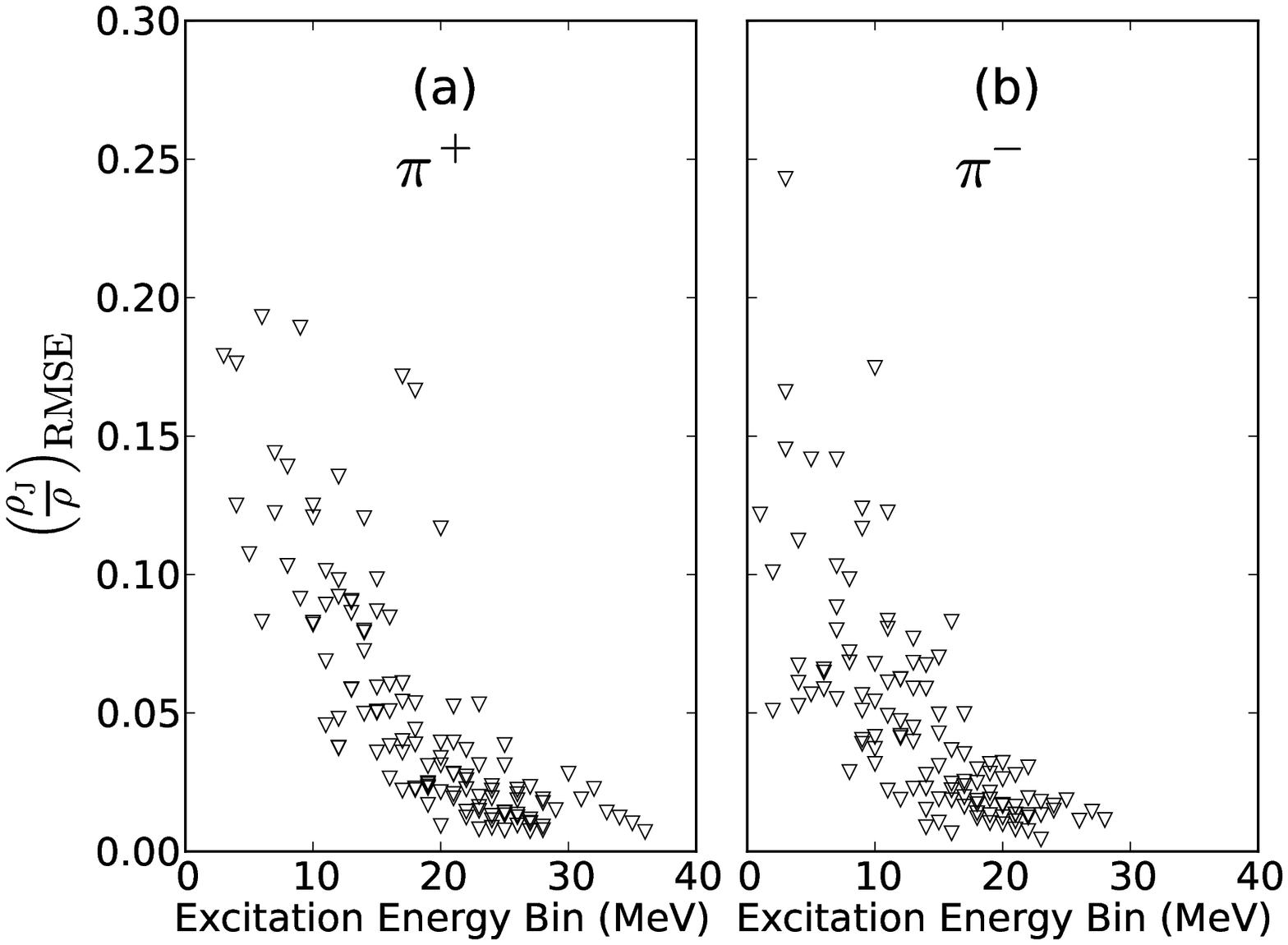}

	\caption{
	$(\rho_{J}/\rho)_{\mathrm{RMSE}}$ (defined in Eq.~\ref{RMSE}) for all analyzed
	psd$_{\frac{5}{2}}$ cross-shell nuclides. 		\label{fig:err_parity}}
\end{figure}

The degree of improvement in $(\rho_{J}/\rho)_{\mathrm{RMSE}}$
appears to be independent of any structural differences in the $sd$ and
$pf$ shell nuclides, as illustrated in Fig.~\ref{fig:err_struc}.  
The cross-shell nuclides in the $p$-$sd_{5/2}$ space, Fig.~\ref{fig:err_parity},  exhibit a similar improvement
in $(\rho_{J}/\rho)_{\mathrm{RMSE}}$, though because the levels are split
between parities the level density is lower than that of the $sd$ and $pf$
shell nuclides, and thus the improvement occurs at higher excitation energies
than with the latter. 

%Solving analytically for the spin cut-off factor by taking
%$\langle J(J+1) \rangle$ over (\ref{eq:ericson}),\\
%% is there a reference for this? 
%\begin{equation}
%	\langle J(J+1) \rangle = \int_0^{\infty}
%	dJ\;J(J+1)\;\frac{\rho_{J}}{\rho},
%\end{equation}
%we derive the following expression for the spin cut-off factor
%($\sigma_{\mathrm{approx}}$):\\
%\begin{equation} \label{eq:approx}
%	\sigma_{\mathrm{approx}}=
%	\sqrt{\frac{1}{2}\langle J(J+1) \rangle+\frac{1}{8}}.
%\end{equation}
%The overall results are nearly identical when applying the same analysis
%pipeline to this approximation versus $\sigma_{\mathrm{fit}}$ as seen
%in the progression of the momentum distribution in Figure \ref{fig:distro} as
%well as Figure %\ref{fig:rmse_approx}.

\begin{figure}
\subfloat{
\includegraphics[width=0.5\textwidth]{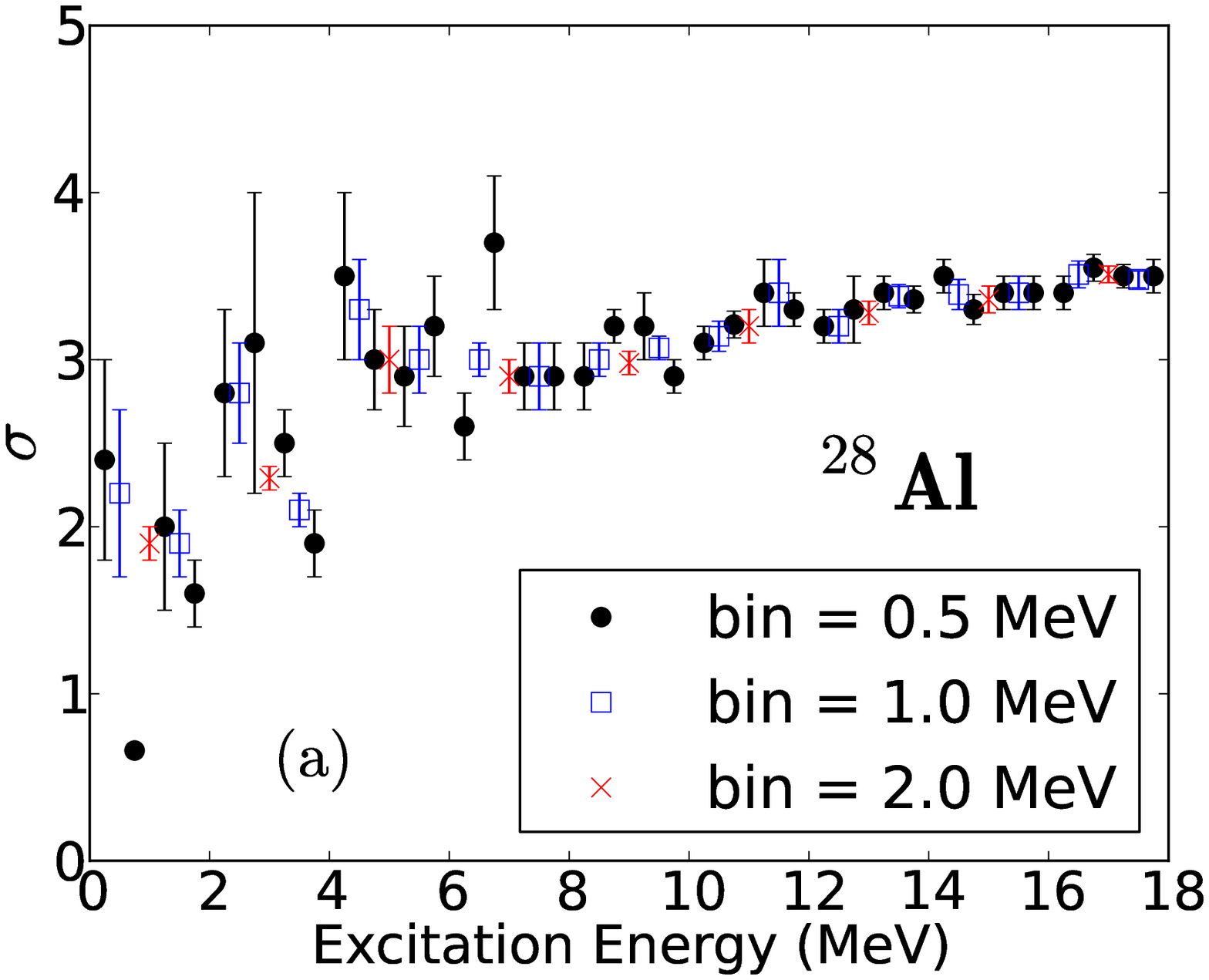}
}
\subfloat{
\includegraphics[width=0.5\textwidth]{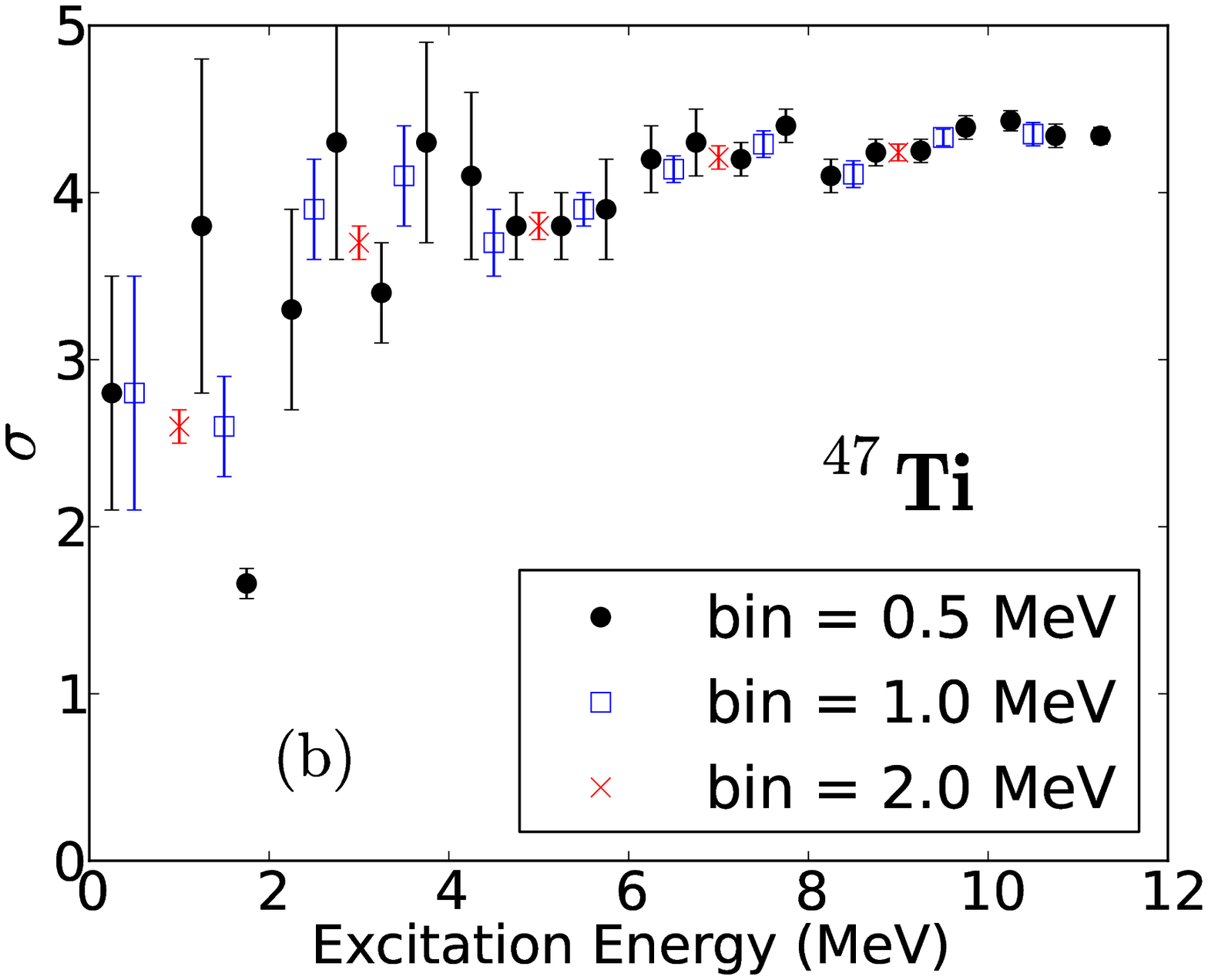}
} 
\caption{(Color online) Dependence of fit spin-cutoff factors on energy bin size 
for $^{28}$Al in the $sd$-shell (left) and $^{47}$Ti in the $pf$-shell.
 } \label{bindependence}
\end{figure}

We also investigated the dependence upon the bin size.  While the above results were for standard bin sizes of 1.0 MeV, 
we also used bins of 0.5 and 2.0 MeV.  The best-fit values were insensitive to the bin size, as 
illustrated in Fig.~\ref{bindependence}, although the error bars depended upon the bin size, which is primarily an effect 
of the number of levels in a bin. 

%Finally: our initial reduced $\chi^2$, before we scaled the errors, was almost always less than 1, 
%often significantly. 
%A reduced $\chi^2 < 1$ often suggests an overfitting, which with just one parameter is improbable 
%in our case. Among possible causes for the small reduced $\chi^2$ are: non-Gaussian errors; 
%correlated errors; and finally approximating a discrete-valued function with a continuous 
%function. Considering only non-Gaussian errors, 
%because of the well-known `level repulsion' in 
%nearest-neighbor distributions in nuclear spectra\cite{LevelSpacing}, the levels are not independent of each 
%other (or, to put it another way, the spectrum for a given $J^\pi$ is more rigid than the assumption
%of Gaussian-distributed errors would suggest). We crudely tested this idea by using 
%$\Delta \rho_J = \rho_J^{1/4}$.  Our initial reduced $\chi^2$ were larger, but overall our best fit values of $\sigma$ did not change and most error bars on 
%$\sigma$ did not change much. 

We therefore conclude that the spin-cutoff factor is a statistically effective parameterization 
of the $J$-dependence of the level density, particularly at high energy.  

As a final note, there is both experimental \cite{vEB08} and theoretical \cite{ALN07} evidence of an odd-even staggering 
relative to the Ericson parameterization, i.e. suppression of odd-$J$ densities and enhancement of even-$J$ densities 
in even-$Z$, even-$N$ nuclides.  We found no consistent evidence for such odd-even staggering in our calculations, for while 
some select energy bins did appear to exhibit it (for example as in Fig. 2), neighboring energy bins showed the opposite trend. 
Our lack of odd-even staggering might be due in part to our modest model space, necessary for 
full-configuration diagonalization, compared to the larger model spaces used in \cite{ALN07}. 

\section{Monte Carlo calculations}

\label{QMC}

We attained our level densities above through laborious diagonalization 
of shell-model Hamiltonians, but such calculations are not practical for large model spaces.  
Therefore we consider other techniques to compute the level density and the spin-cutoff factor, in 
particular shell-model Monte Carlo (SMMC) \cite{Lang93,KDL97,ALN07, VRHA09, Orm97,NA97,NA98,LA01,ABFL05,AFN08}. 
In this section, unlike the previous sections, we use the state density, which includes the $2J+1$ degeneracy in $M_J$, 
because it arises more naturally for 
SMMC. The normalized Ericson function for the state density is 
\begin{equation}
f_\mathrm{state}(J) = \frac{2J+1}{\sqrt{ 8 \pi} \sigma ^3} \exp(-J(J+1)/2\sigma^2).
\label{Ericson_state}
\end{equation}

Most theoretical calculations of the density of states start from thermodynamics, 
as  the partition function is the Laplace transform of the density of states:
\begin{equation}
{\cal Z}(\beta) =  \mathrm{tr} \, e^{-\beta \hat{H}} =\int e^{-\beta E} \rho_\mathrm{state}(E) dE
= \sum_i (2J_i+1) \exp(-\beta E_i),
\end{equation}
where the trace and the sum are over many-body states and one can interpret $\beta$ as inverse temperature.  The density 
of states itself can be written as $\rho_\mathrm{state}(E) = \sum_i (2J_i+1) \delta(E-E_i)$.  The Laplace transform can be inverted with 
reasonable reliability through the saddle-point approximation \cite{Boh69}, 
and so now one has the problem of computing the nuclear partition function.

One avenue to compute the many-body partition function is to use auxiliary-field path integrals 
 to evaluate $\exp( -\beta \hat{H})$.  While such an approach has been successful, it requires a nontrivial 
amount of computing time; furthermore because one must evaluate the resulting integrals through Monte Carlo sampling, one has to 
use a well-posed interaction that is free or mostly free of the sign problem \cite{Lang93,KDL97,ALN07}.

\begin{figure}
%	\centering
\subfloat{
	\includegraphics[width=0.5\textwidth]{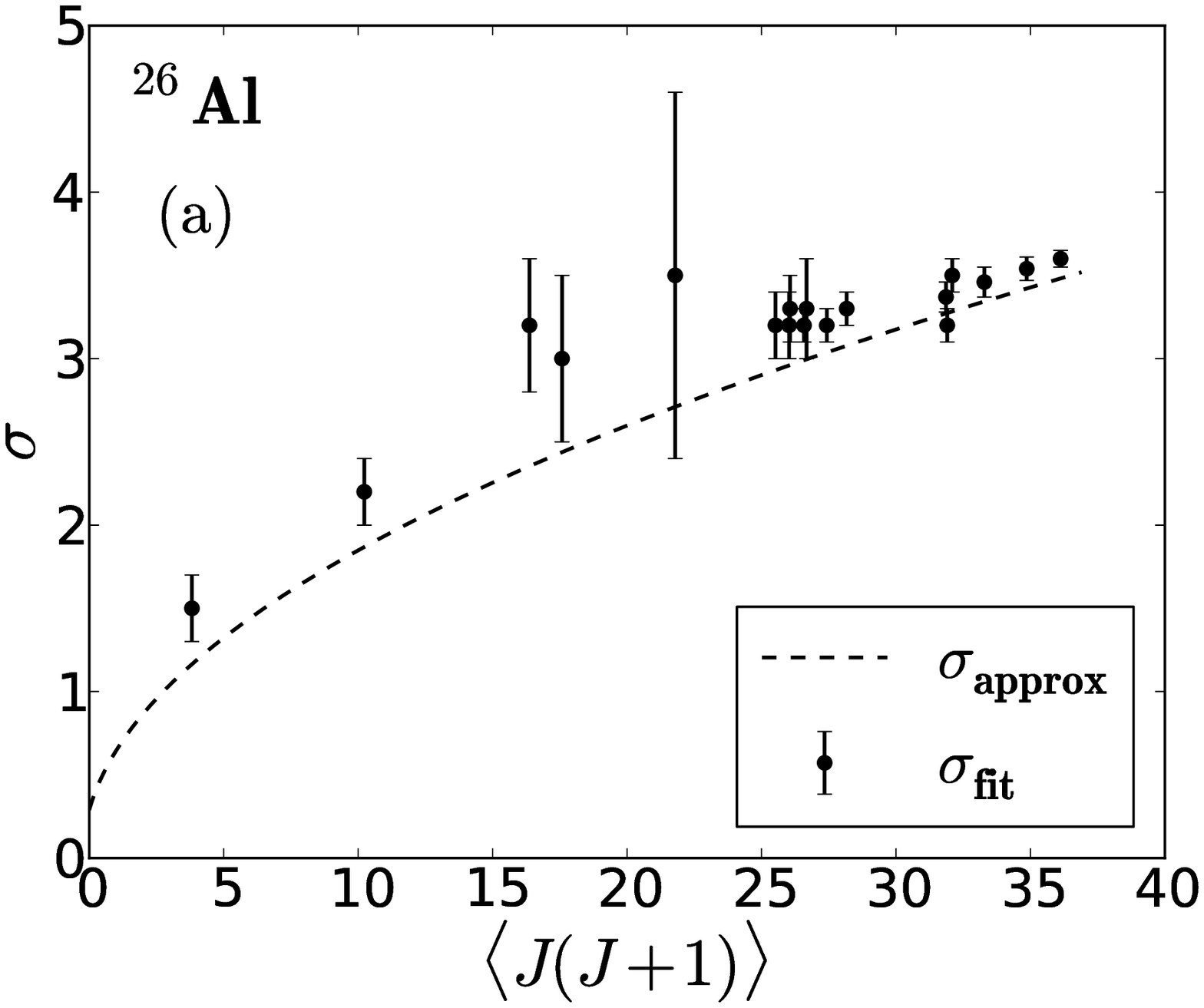}
}
\subfloat{
	\includegraphics[width=0.5\textwidth]{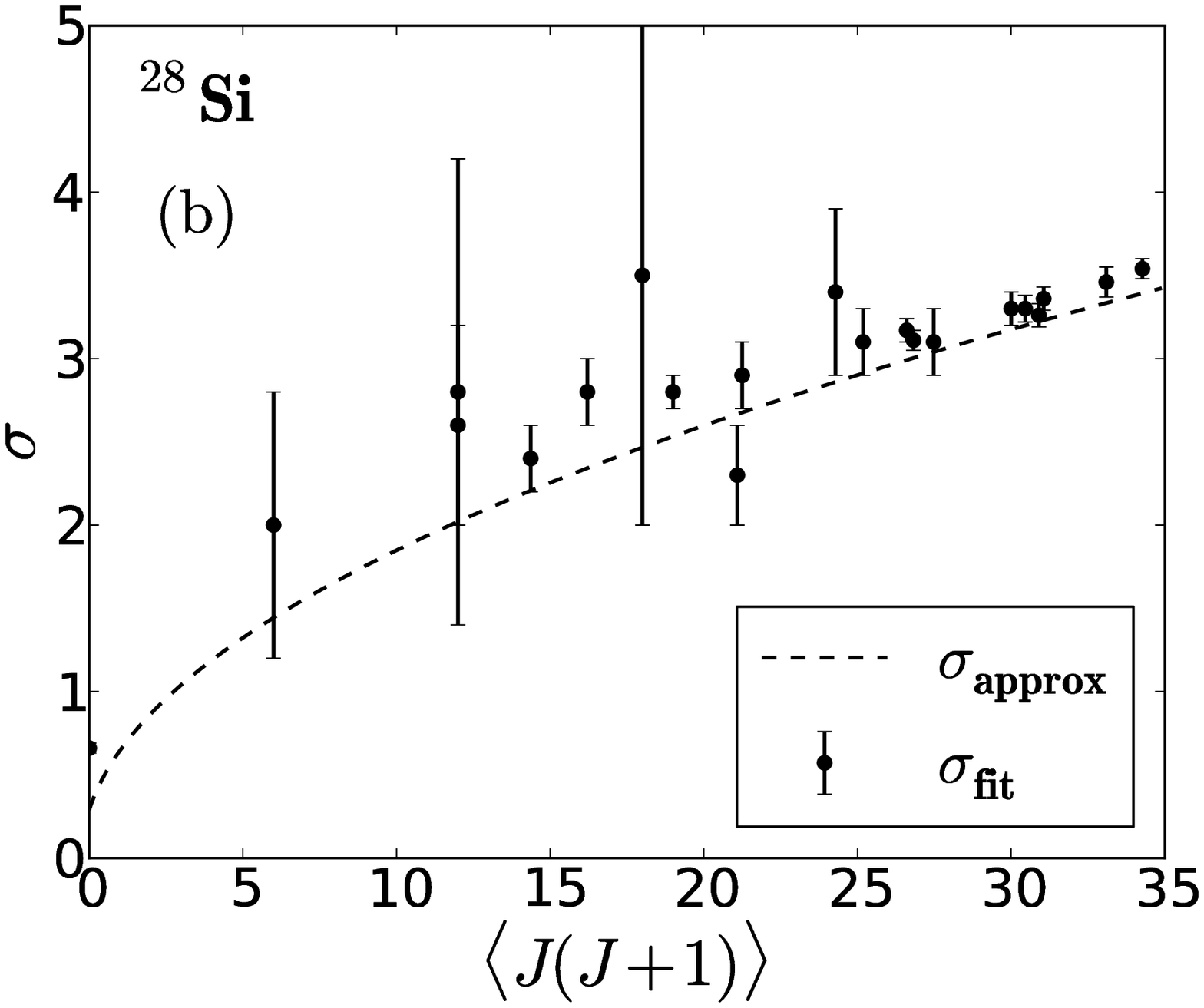}
}  \\
\subfloat{
	\includegraphics[width=0.5\textwidth]{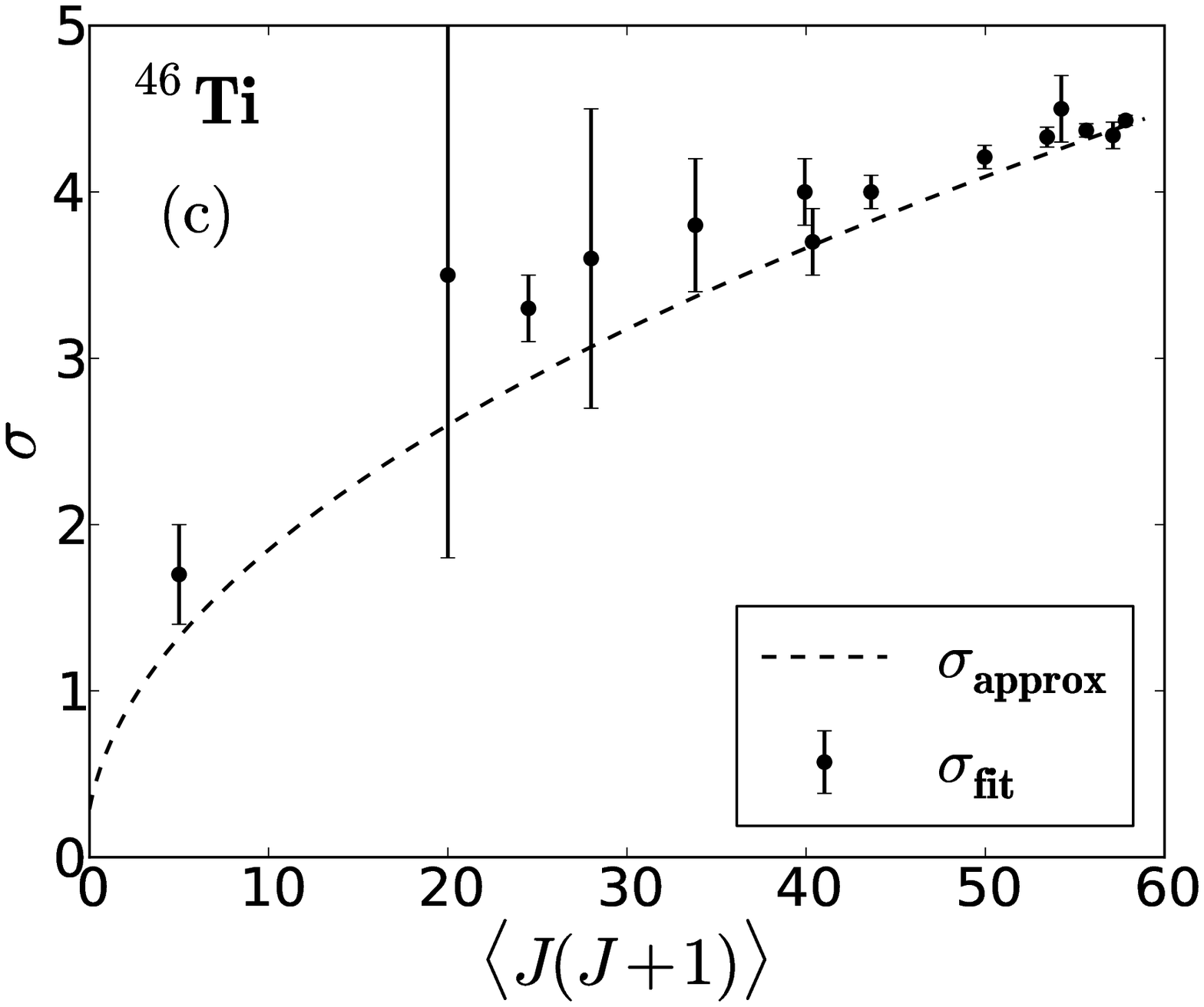}
}
\subfloat{
	\includegraphics[width=0.5\textwidth]{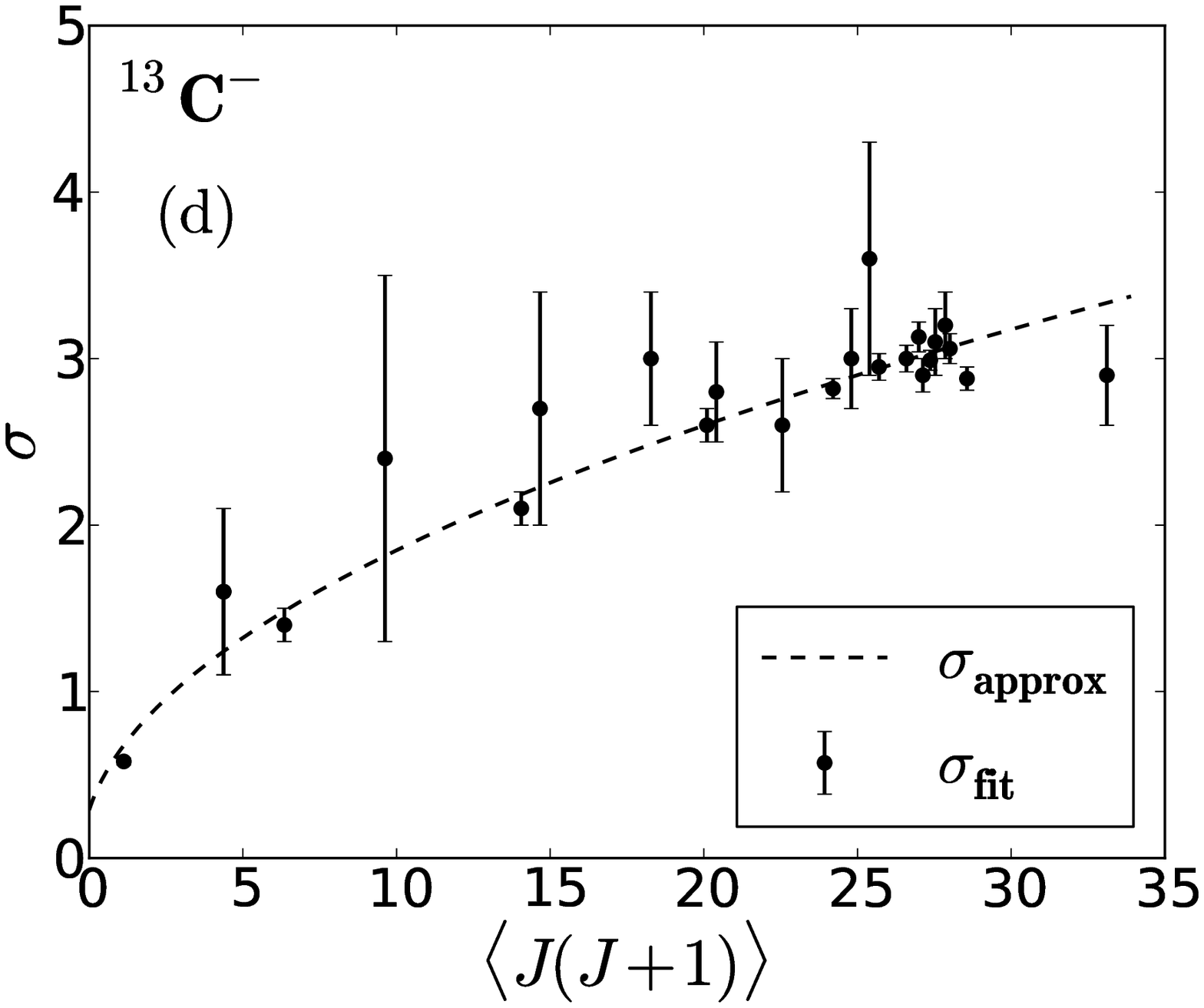}
}

	\caption{Comparison of the spin-cutoff factor extracted by least-squares-fit 
($\sigma_\mathrm{fit}$) against the spin-cutoff factor extracted from 
	 the average value of $J(J+1)$ in an energy bin ($\sigma_\mathrm{approx}$), 
for several representative nuclides.}\label{fig:scf_approx}
\end{figure}

The spin-cutoff factor can be related directly to the average value of $J(J+1)$ in an energy bin. 
 One can either assume  the nucleon spins add together randomly \cite{ABFL05} or simply  replace the discrete Ericson 
function (\ref{Ericson_state}) by a continuous function of $J$ and integrate.  Then one finds \cite{ABFL05,Qur03}
\begin{equation}
3\sigma^2 \approx \langle J(J+1) \rangle +\frac{1}{4} \label{sigapprox}
\end{equation}
where $\langle \ldots \rangle$ indicates an average over the state density, that is, 
\begin{equation}
\langle J(J+1) \rangle
= \frac{ \sum_{i \in \Delta E}   J_i (J_i+1)  (2J_i+1)}
{ \sum_{i \in \Delta E}  (2J_i + 1)}
\end{equation}
where $i \in \Delta E$ refers to a sum over states in an energy bin $\Delta E$. 
 Using our densities, we found this approximation worked well, as illustrated in 
Fig.~\ref{fig:scf_approx}, although our fit $\sigma$ tended to be a little higher than 
the $\sigma_\mathrm{approx}$ found using (\ref{sigapprox}). If we used the level density rather than the 
state density we got a better fit, but the state density is natural for SMMC calculations.

The authors of \cite{ABFL05} relate the spin-cutoff factor to the moment of inertia by way of the 
spin response function in imaginary time. In fact, in the SMMC it is straightforward to compute the thermal average of the 
expectation value of the angular momentum,
\begin{equation}
\langle \hat{J}^2(\beta)  \rangle= \frac{ \mathrm{tr} \, \left [ \exp(-\beta \hat{H} )\hat{J}^2 \right ]}{ {\cal Z}(\beta)}
= {\cal Z}(\beta)^{-1} \sum_i \exp(-\beta E_i)(2J_i+1)  J_i (J_i+1). \label{thermavg}
\end{equation}
This is just the Laplace transform of the sum of $J(J+1)$ in an energy bin, that is
if we define the average as
\begin{equation}
\langle J(J+1) (E)\rangle = \frac{ \sum_{i }(2J_i+1) J_i(J_i+1) \delta(E-E_i) }{ \rho_\mathrm{state}(E)}
\end{equation}
(we deliberately use the notation $\hat{J}^2$ for the thermal average and $J(J+1)$ for the 
energy average, to help distinguish the two notationally) 
then
\begin{equation}
{\cal Z}(\beta) \langle \hat{J}^2(\beta)  \rangle = \int e^{-\beta E} \rho_\mathrm{state}(E) \langle J(J+1)  (E) \rangle dE.
\end{equation}

In the same way one can invert the Laplace transform via the method of steepest descent, one can 
also extract $\langle J(J+1)  (E) \rangle$, or rather, extract $\rho(E) \langle J(J+1)  (E) \rangle$ and
then divide by the level density, as follows. First, find the saddle point, which yields the effective temperature 
for a given target excitation energy:
\begin{equation}
E = \left . - \frac{\partial \ln {\cal Z}(\beta) + \ln \langle \hat{J}^2  (\beta) \rangle }{\partial \beta}
\right |_{\beta = \beta_\mathrm{eff} }.
\end{equation}
For a given effective temperature $T_\mathrm{eff}= 1/\beta_\mathrm{eff}$, the energy $E$ is different 
from that for the saddle point condition for the level density, because of the additional term.

\begin{figure}
\includegraphics[width=7cm]{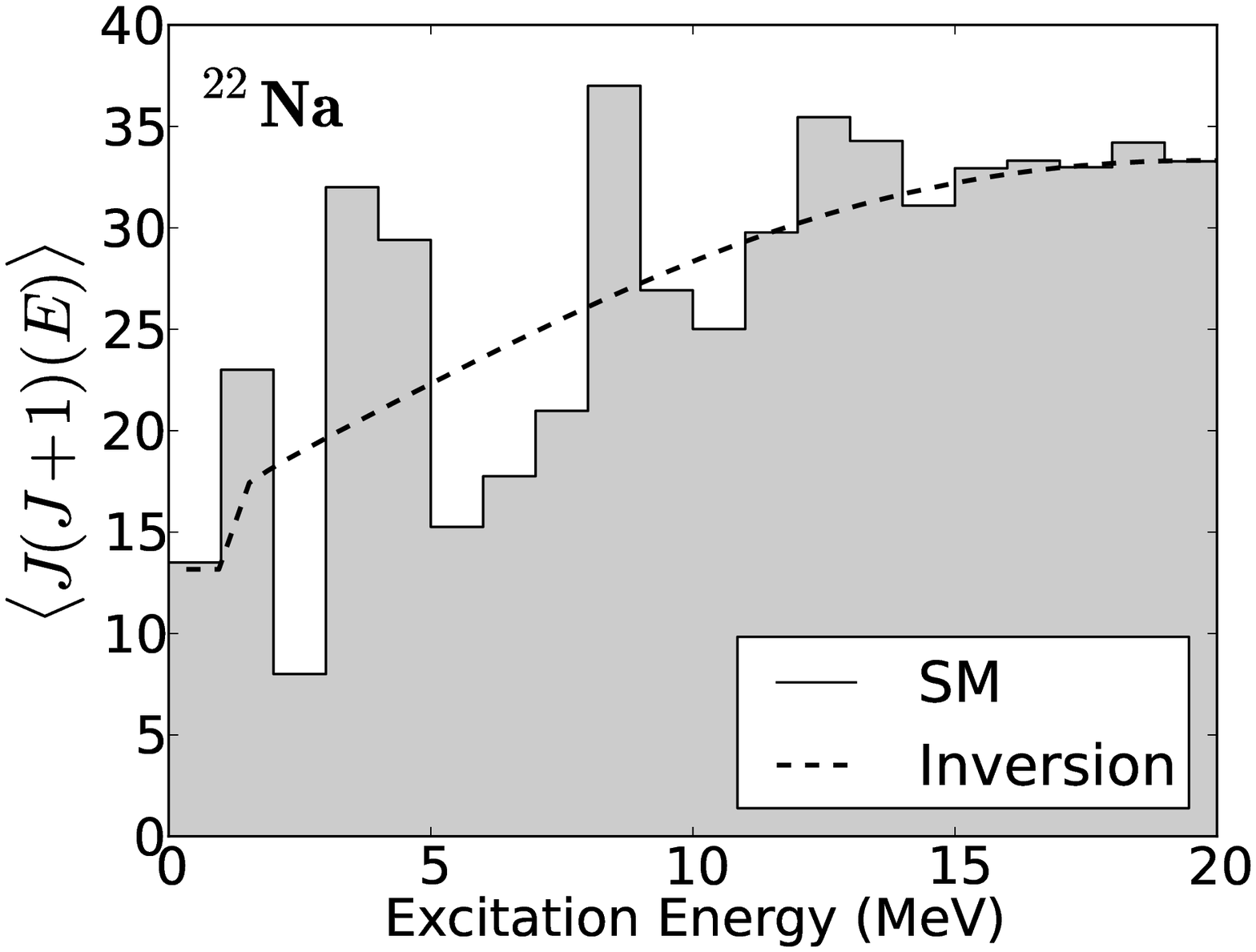}
\caption{Comparison of the average value of $J(J+1)$ in 1 MeV energy bins, calculated 
directly from shell-model diagonalization spectrum for $^{22}$Na in the $sd$-shell (`SM', histogram) and from inverting 
the thermal average, Eq.~(\ref{thermavg}), using Eq.(\ref{invertJ2}) (`Inversion', dashed line).
}
\label{invertMC_Na22}
\end{figure}

\begin{figure}
\subfloat{
\includegraphics[width=7cm]{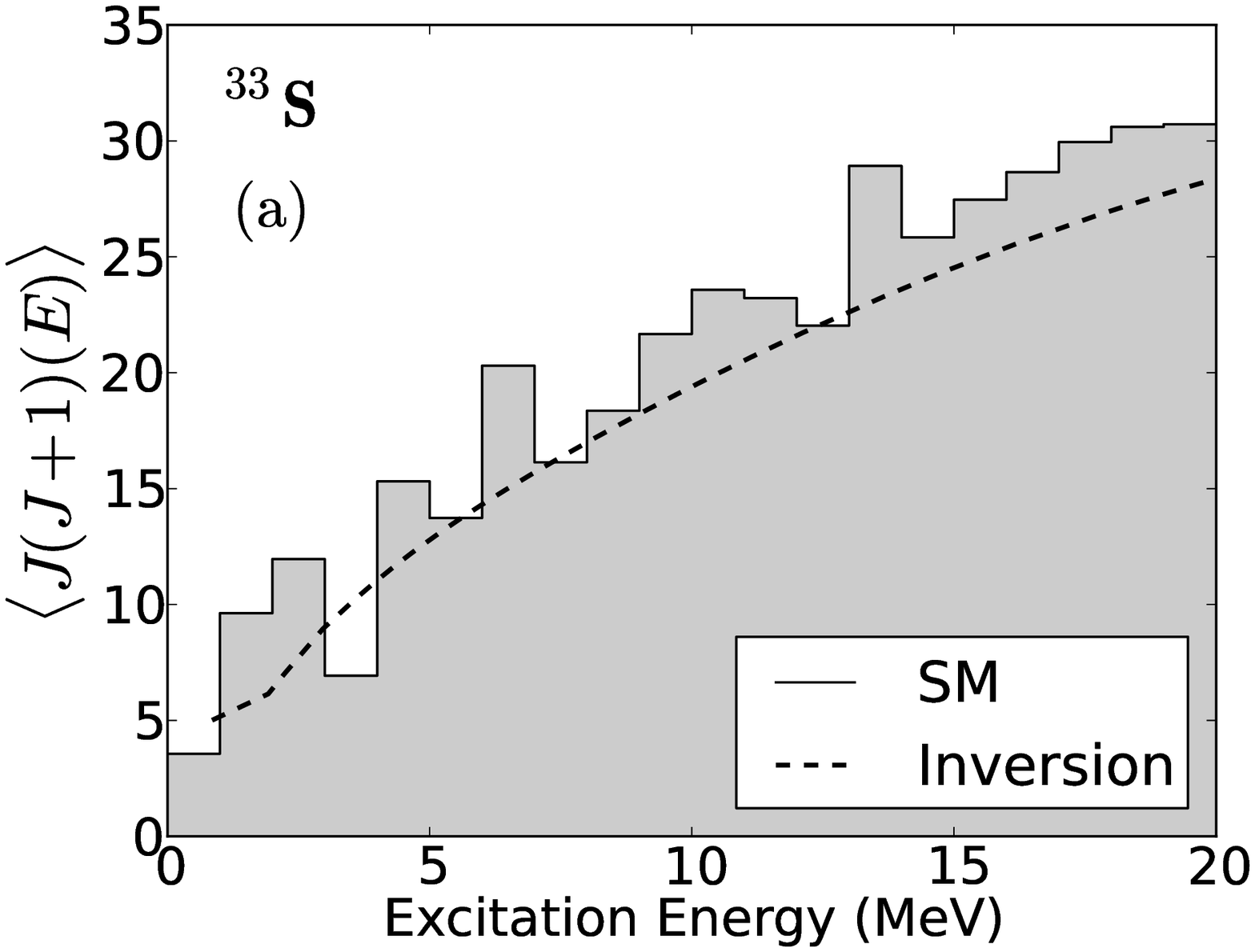}
}
\subfloat{
\includegraphics[width=7cm]{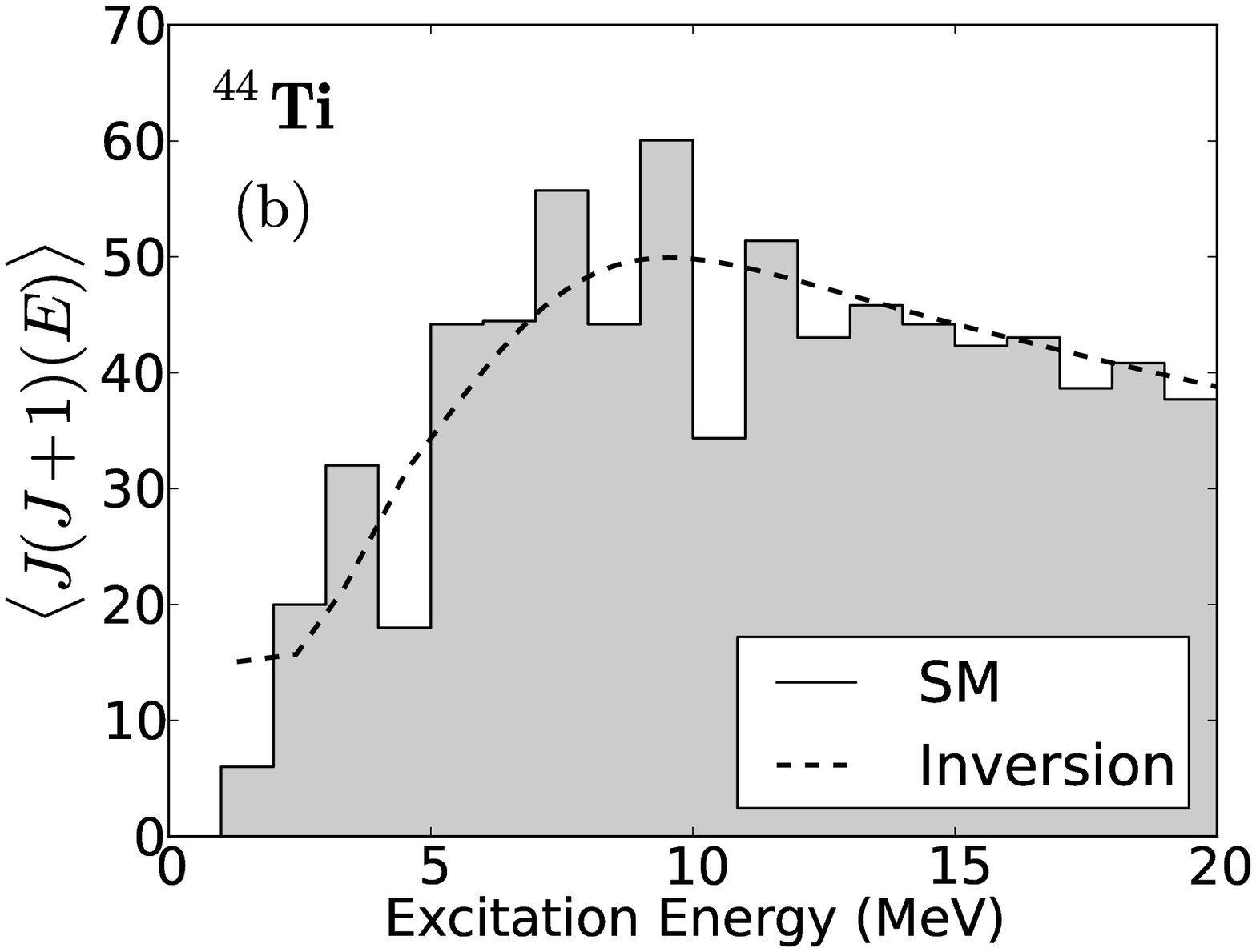}
}
\caption{Same as Fig.~\ref{invertMC_Na22} but for $^{33}$S in the $sd$-shell and 
$^{44}$Ti in the $pf$-shell (right). The turnover for $^{44}$Ti is due to the finite model space; 
the other nuclides also display a turnover, but at higher excitation energies. \label{invertMC}
}

\end{figure}

One must also compute the second derivative:
\begin{equation}
C =  \left . \frac{\partial^2 \ln {\cal Z}(\beta) + \ln \langle \hat{J}^2  (\beta) \rangle }{\partial \beta^2}
\right |_{\beta = \beta_\mathrm{eff} }.
\end{equation}
The final result of the method of steepest descent is
\begin{equation}
\rho_\mathrm{state}(E) \langle J(J+1)  (E) \rangle = \frac{{\cal Z}(\beta_\mathrm{eff})  \langle \hat{J}^2  (\beta_\mathrm{eff}) \rangle 
\exp(\beta_\mathrm{eff} E) }{\sqrt{ 2\pi C}}. \label{invertJ2}
\end{equation}
The state density itself is computed using the exact same kind of inversion, only without the $ \langle \hat{J}^2  (\beta) \rangle$ term in the derivatives.   Although 
 we did not carry out any Monte Carlo calculations, we modeled such a calculation by computing several 
exact and complete spectra, namely $^{22}$Na and $^{33}$S in the $sd$ shell and $^{44}$Ti in 
the $pf$ shell, each of which have dimensions of 4000-7000; from those spectra 
we constructed the partition 
function as well as the thermal average of $J(J+1)$.  We then inverted using Eq.~(\ref{invertJ2}) 
and found the average value of $J(J+1)$ in an energy bin is well-reconstructed, 
as shown in Fig.~\ref{invertMC_Na22},\ref{invertMC}. 
(Note: although in nature one expects the average $J(J+1)$ to increase with excitation energy,
with finite model spaces it must turn over at some point, as seen here for $^{44}$Ti.)

Although a practical $J$-projection scheme for level densities has been implemented \cite{ALN07}, 
by computing an $M$-projected density and taking $\rho_J = \rho_{M=J} - \rho_{M=J-1}$, 
simply computing $\langle \hat{J}^2 (\beta) \rangle $ is significantly computationally cheaper, and appears 
to be effective in arriving at the spin-cutoff factor. 

\section{Conclusions and acknowledgements} 

We have carried out a detailed investigation into the spin-cutoff parameterization Eq.~(\ref{Ericson}) 
of the $J$-dependence 
of the level density, with particular attention paid to the error bars in the fit to $\sigma$.   We confirmed that 
the spin-cutoff parameterization is a good one, and that as the statistics improve the error bars on $\sigma$ 
decrease. There was no qualitative difference in behavior for different nuclides, and the best fit values were insensitive 
to bin size.    

We also showed that a simple energy average, $\langle J(J+1) (E_x) \rangle$ gives a good value for the spin-cutoff parameter 
and that one can extract the energy average from a thermal average $\langle \hat{J}^2(\beta) \rangle$ which one 
might get naturally out of a quantum Monte Carlo calculation.  Such an approach would be computationally more 
efficient than directly calculating $J$-projected densities, although the latter are achieveable.

The U.S.~Department of Energy supported this investigation through
grant DE-FG02-96ER40985.


\begin{thebibliography}{99}
\bibitem{SJsuppl} See Supplmental Material at [URL will be inserted by publisher] for graphs for all nuclides listed 
in Table \ref{tab:list}.
\end{thebibliography}
\end{document}